\documentclass[acmsmall]{acmart}
\AtBeginDocument{%
  }

\setcopyright{cc}
\setcctype{by}
\acmJournal{POMACS}
\acmYear{2025} \acmVolume{9} \acmNumber{1} 
\acmArticle{4} \acmMonth{3} 
\acmPrice{}
\acmDOI{10.1145/3711697}

\usepackage{graphicx}
\usepackage[]{hyperref} 
\hypersetup{
    colorlinks=true,urlcolor=purple,linkcolor=cyan,citecolor=cyan
}
\usepackage{amsmath}
\usepackage{booktabs}
\usepackage[labelformat=simple]{subcaption}

\usepackage{xcolor}

\newif\ifdraft
\draftfalse 

\ifdraft
        \newcommand{\ttsuchiy}[1]{\textcolor{blue}{[[Taro: #1]]}}
        \newcommand{\nicolasc}[1]{\textcolor{brown}{[[Nicolas: #1]]}}
        \newcommand{\lzhou}[1]{\textcolor{red}{[[Liyi: #1]]}}
        \newcommand{\kaihuaq}[1]{\textcolor{purple}{[[Kaihua: #1]]}}
        \newcommand{\arthur}[1]{\textcolor{green}{[[Arthur: #1]]}}

\else
        \newcommand{\ttsuchiy}[1]{}
        \newcommand{\nicolasc}[1]{}
        \newcommand{\lzhou}[1]{}
        \newcommand{\kaihuaq}[1]{}
        \newcommand{\arthur}[1]{}

\begin{document}
%
\title{Blockchain Amplification Attack}


\author{Taro Tsuchiya}
\email{ttsuchiy@andrew.cmu.edu}
\affiliation{%
  \institution{Carnegie Mellon University}
  \city{Pittsburh}
  \country{USA}
}
\author{Liyi Zhou}
\email{liyi.zhou@sydney.edu.au}
\affiliation{%
  \institution{University of Sydney, Decentralized Intelligence AG, Berkeley RDI}
  \city{Sydney}
  \country{Australia}
}
\author{Kaihua Qin}
\email{kaihua@qin.ac}
\affiliation{%
  \institution{Yale University, Decentralized Intelligence AG, Berkeley RDI}
  \city{New Haven}
  \country{USA}
}
\author{Arthur Gervais}
\email{arthur@gervais.cc}
\affiliation{%
  \institution{University College London, Decentralized Intelligence AG, Berkeley RDI}
  \city{London}
  \country{UK}
}
\author{Nicolas Christin}
\email{nicolasc@andrew.cmu.edu}
\affiliation{%
  \institution{Carnegie Mellon University}
  \city{Pittsburgh}
  \country{USA}
}

\renewcommand{\shortauthors}{Taro Tsuchiya, Liyi Zhou, Kaihua Qin, Arthur Gervais, and Nicolas Christin}

\begin{abstract}
Strategies related to the blockchain concept of 
Extractable Value (MEV/BEV), such as arbitrage, front-, or 
back-running create strong economic incentives for network nodes
to reduce latency.
\emph{Modified nodes}, that minimize transaction validation time and neglect to filter invalid transactions in the Ethereum peer-to-peer (P2P) network, introduce a novel attack vector---a \textit{Blockchain Amplification Attack}.
An attacker can exploit those modified nodes to amplify invalid transactions thousands of times, posing a security threat to the entire network. 
To illustrate attack feasibility and practicality in the current Ethereum network (``mainnet''), we 1) identify thousands of similar attacks in the wild, 2) mathematically model the propagation mechanism, 3) empirically measure model parameters from our monitoring nodes, and 4) compare the performance with other existing Denial-of-Service attacks through local simulation. 
We show that an attacker can 
amplify network traffic at modified nodes by a factor of 3,600, and cause
economic damages of approximately 13,800 times the amount needed to carry out the attack. 
Despite these risks, aggressive latency reduction may still be profitable 
enough for various providers to justify the existence of modified nodes. 
To assess this trade-off, we 1) simulate the transaction validation process in a local network and 2) empirically measure the latency reduction by deploying our modified node in the Ethereum test network (``testnet''). 
We conclude with a cost-benefit analysis of skipping validation and provide mitigation strategies against the blockchain amplification attack.
\end{abstract}

\ccsdesc[500]{Security and privacy~Denial-of-service attacks}
\ccsdesc[500]{Networks~Network measurement}


\keywords{Blockchain, Security, DoS, P2P network, Ethereum}

\received{October 2024}
\received[revised]{January 2025}
\received[accepted]{January 2025}

\maketitle

\section{Introduction}
\label{sec:intro}
In ``Flash Boys'' \cite{lewis2014flash}, Lewis famously describes how,
in the traditional stock market, lower network latency to stock
exchange matching engines provides such a competitive advantage that
trading firms have started to physically shorten network links -- using,
e.g., dedicated fiber -- and routinely optimize network configuration to
reduce latency.

Could the same type of optimizations apply to decentralized             
finance (DeFi), i.e., in the context of blockchain peer-to-peer (P2P)         
networks? In principle, yes: the concept of Extractable Value           
(MEV/BEV)~\cite{daian2020flash, qin2022quantifying}, i.e., maximizing   
profit by reordering transactions, implies that centralized business    
entities have an incentive to reduce latency    
to deliver pending transactions. Being the fastest to get transactions  
included in a block -- even if just by milliseconds -- 
can attract users, bots, and validators who         
can profit from arbitrage, front- or back-running, and sandwich         
attacks~\cite{zhou2021high}.

\begin{figure}[]
    \centering
    \includegraphics[width=0.85\linewidth]{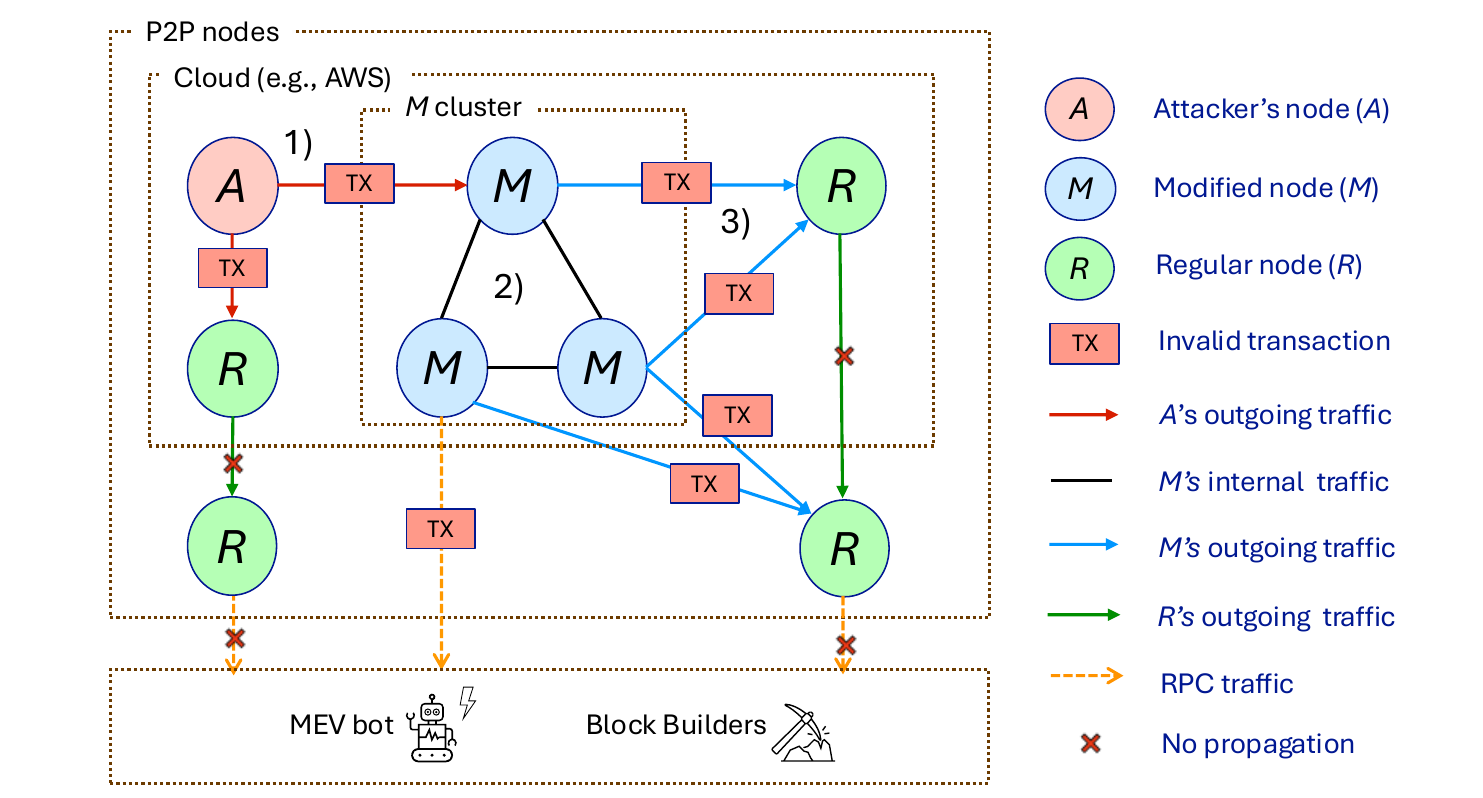}
    \caption{Overview of the Blockchain Amplification Attack, illustrating how one invalid transaction amplifies in the network.}
    \label{fig:edos_sketch}
\end{figure}

As a case in point, we recently notice that certain \emph{modified     
nodes} in the Ethereum network tailor their configurations, sidestep transaction validation     
processes to shorten latency, which introduce invalid transactions that   
are not supposed to propagate in the network.                             
Unfortunately, these optimizations open the door to a \emph{Blockchain Amplification Attack}, which we 
formalize and study in this paper. 
In this attack, depicted in Figure~\ref{fig:edos_sketch}, 1) the adversary $A$ sends an invalid transaction (denoted by the red arrow) to a modified node $M$, that 2) shares it with other modified nodes (black arrows), which will 3) collectively further propagate it (blue arrows) to the rest of the network, amplifying the effect of the invalid transaction on the network.  
Indeed, the attacker can not only hamper modified nodes (and cause economic damage through excessive traffic), but, by also spread invalid transactions to slow down the \textit{entire} network. 
This paper fully describes such attack opportunities, documents the prevalence of similar attacks in the wild, models network damage empirically and by simulation, and quantifies the trade-offs associated with skipping validations.
Due to the absence of a penalty mechanism for sending invalid transactions, we demonstrate that the attack is feasible, impactful, and cost-effective in the current Ethereum P2P network (``mainnet''). 

We first mathematically model the transaction propagation mechanism and pinpoint the parameters that affect how an attacker's invalid transactions can get amplified in the whole network (i.e., the \textit{amplification factor}). 
While considering the specifics of each software/version on transaction forwarding, our model establishes a framework to quantify the increase in both the amount and the economic cost of invalid transactions.

We next demonstrate the attack feasibility in the wild.
In particular, we analyze the set of pending transactions (usually referred to as ``\emph{txpool}'') for centralized entities that deliver transactions to users. 
Some propagate transactions much faster than the others but propagate invalid transactions, suggesting the lack of validation checks. 
We identify 2,591 similar attack instances across 345 Ethereum 
addresses that target these entities.  
We classify those instances based on the attacker's observed strategies and estimate the attack cost, size, and intensity.  

To correctly estimate amplification factors (i.e., the impact of the attack) in the current mainnet, we design two customized monitoring Geth nodes (the most popular Ethereum node client) to infer the network topology. 
Our custom nodes scan the activities on the P2P layer and record specific types of messages \textit{before} they undergo processing, resulting 2.5 billion observations. 
Our infrastructure allows us to perform a new method for inferring the number of \textit{active} peer connections---a cost-effective and ethical alternative to previously proposed methods~\cite{delgado2019txprobe, daniel2019map, cao2020exploring, li2021toposhot}. 
We further show that 1.5\% of nodes exhibit lenient transaction validation (``\emph{modified nodes}''). 
Looking at the \texttt{git commit} hash of the modified nodes, which shows unique and non-public properties of the software, many of these nodes appear to originate from the same entity. 
All in all, 
we show that the attacker can 
amplify outgoing network traffic at modified nodes by a factor of \textit{at least} 3,600, and cause economic damages 13,800 times the amount needed to carry out the attack. 

In addition, we conduct attack simulations in a local network and confirm that our proposed attack can evict as many honest transactions (from both the txpool and the block) as other proposed DoS attacks but at significantly lower costs.

Furthermore, we experiment to quantify the benefits of skipping transaction validation, specifically the amount of time saved for each validation process. 
We first simulate the process of Geth's transaction validation in a locally controlled environment. 
We find that validating one transaction takes roughly 1 millisecond and checking account status (nonce/balance) is the most time-consuming (86\% of total validation time). 
We also empirically measure the latency reduction by deploying the modified/regular nodes in the Ethereum test network (``testnet'') for 14 weeks. 

Based on the latency gains from skipping validations, we estimate the maximum economic gains (0.00002 ETH $\approx$ \$0.050 per millisecond~\cite{wahrstatter2023time, schwarz2023time,babel2024prof}), and compare it with the potential attack damage. 
We finally offer three possible mitigations: 1) enforcing a stricter txpool policy, 2) postponing the validation process, and 3) introducing a node reputation system.

The contributions of this paper are as follows.
\begin{itemize}
    \item We mathematically define a Blockchain Amplification Attack and show attack feasibility based on similar attacks found in the wild.
    \item By designing and deploying our custom monitoring nodes, we infer the Ethereum P2P network topology and illustrate the efficiency of the attack.
    \item Simulations on the local network confirm that 
    the blockchain amplification attack is more cost-effective for the attacker than other existing DoS attacks. 
    \item We simulate and empirically measure the transaction validation process for latency reduction. 
\end{itemize}

\section{Background}
\label{sec:bg}
We next present relevant background on blockchain peer-to-peer networks, and associated security issues.
\subsection{Blockchain and P2P network}
Three aspects of the blockchain ecosystem are particularly pertinent
to our paper: 1) P2P network formation, 2) transaction 
validation, and 3) interactions with users outside
the P2P network.

Ethereum nodes use the 
Kademlia~\cite{maymounkov2002kademlia} distributed hash table for connectivity. 
When a node launches, it initially connects to hardcoded bootstrap nodes, which provide a list of potential peers.
The node identifier, known as ``enode,''  comprises the node ID (randomly generated from the node's public key), IP address, and port number. 
The node calculates the distance between possible peers based on node IDs, divides them into buckets, and populates the list of peers,\footnote{The node calculates the XOR of two node IDs: Keccak256(node1's id) XOR Keccak256(node2's id) as a distance, and group the possible peers into 256 buckets: for each $i$ (bucket), where $2^{i} \leq distance < 2^{i+1}$)} making peer selection arbitrary~\cite{ethereum_discv4}.
Geth by default allows 50 connections, while Erigon and Nethermind permit 100.
Geth allocates one-third of connections to active peer searching and uses the other two-thirds for passively accepting inbound connections. 
Ethereum's mainnet shares the same underlying P2P network with other chains (e.g., Ethereum Classic or Testnets such as Holesky). 
Geth nodes immediately disconnect from peers on different chains.

After establishing a connection, nodes initiate the exchange of messages to synchronize blockchain information (blocks and transactions), following the Ethereum Wire Protocol~\cite{ethereum_wire}. 
For every transaction, a node must validate the integrity of each transaction, which contains several parameters, regardless of encoding type.
Ethereum is an account-based ledger, meaning each address maintains the balance and the transaction index called a ``nonce'' (starting from 0, incrementing per transaction). 
The amount of ``gas'' represents the computational costs of the transaction.
We multiply by the gas price to obtain the total transaction fees. 
The transaction is either to 1) send ``value'' to another address, or 2) execute a smart contract where ``data field'' is used to specify the function and its input. 
The node checks the following conditions for each transaction:
\begin{itemize}
    \item The sender has a balance of more than $\text{gas-limit} \times \text{gas-price} + \text{transfer value}$
    \item The sender's nonce is equal to/larger than the current nonce (i.e., greater than the past nonce)\footnote{The acceptable nonce gap depends on each client's implementation}. 
    \item The transaction size is below a pre-defined limit (e.g., 128KB). 
\end{itemize}

Beyond protocol-level verifications, each software incorporates supplementary checks to mitigate DoS (Denial-of-Service) attacks. 
For example, Geth employs the ``gas bump rule,'' requiring the gas fee for a transaction with the same sender and nonce to be higher than the original transaction with a specified increase (e.g., 10\% in Geth).
The node should not accept replicated transactions with the same address, nonce, and gas price.
Li et al.~\cite[\S5.1]{li2021toposhot} or Yaish et al. ~\cite[\S4.1]{yaish2023speculative} provide  detailed explanations. 

If the transaction is valid, it gets added to the buffer referred to as ``txpool'' (or mempool, tx-queue); otherwise, it should be 
dropped. 
Each software implementation imposes its own limit on the number of transactions accepted in the txpool; even legitimate transactions may be discarded if the txpool reaches its maximum capacity. 

When a new transaction enters the node's txpool, the node broadcasts it to the rest of the network using two message options: 1) broadcast \texttt{Transactions} (\texttt{0x02}) -- RLP (Recursive-Length Prefix)-encoded raw transactions that include all parameters, or 2) announce \texttt{NewPooledTransactionHashes} (\texttt{0x08}) -- transmitting only transaction hashes.
Typically, the node broadcasts \texttt{0x02} to a small subset of nodes for efficiency, and announces \texttt{0x08} to the remaining nodes. 
If the node lacks information about a transaction based on its hash (\texttt{0x08}), it requests a peer to send transaction content via \texttt{GetPooledTransactions} (\texttt{0x09}), and the recipient node responds with \texttt{PooledTransactions} (\texttt{0x0a}).
We describe software-specific implementations in \S\ref{subsec:network_waste}.

A validator selects the set of transactions from its txpool and assembles the block. 
However, some transactions, despite passing the pre-check and being included in the txpool, never make it to the chain; these are referred to as \textit{dropped transactions}. 
If the gas price is insufficient, validators may overlook these transactions. 
Subsequently, senders update the gas price, replacing the old transactions, which results in dropped transactions. 
In this paper, we assume transactions are ``dropped'' if they fail to be included in the blockchain within a span of more than seven days from their initial appearance in txpool. 
In Appendix~\ref{appendix:dropped_tx}, we experiment to validate the use of a 7-day blockchain lookup.
Some transactions are sent with insufficient funds or past nonce, making them ineligible for inclusion in the chain. 
We specifically use the term \textit{invalid transaction} for those. 

When users fetch the blockchain data or initiate a transaction, they typically interact with a P2P node through Remote Procedure Call (RPC) requests, and nodes execute these requests.
Given the high cost of maintaining own node, there exist centralized entities (e.g., Infura, Alchemy) that open up their RPC endpoints or direct node connections for users.
Some (e.g., bloXroute, Eden, Chainbound) optimize the network configurations and deliver pending transactions faster than other nodes as a service. 
For example, bloXroute deploys nodes across globe, internally indexes transactions, propagates without block validations (``cut-through routing''), and employs dynamic routing~\cite{klarman2018bloxroute, blx_bdn}.
Users sometimes pay subscription fees to connect to those special nodes. 


\subsection{Security in P2P network}
While a P2P network serves as the foundation of decentralized blockchains, sharing the same information among all peers is challenging, and often results in network forks~\cite{decker2013information}. 
Forks expose the network to double-spending~\cite{karame2012double, gervais2016security}, eclipse~\cite{heilman2015eclipse} or selfish mining~\cite{eyal2018majority} attacks that all 
partition the network to block information flow.
In the PoW (Proof of Work) blockchain, Luu et al.~\cite{luu2015demystifying} proposed the ``Verifier's dilemma,'' wherein rational miners opt to forgo \textit{block} validations to allocate more time to PoW mining.  
Das et al.\cite{das2021tuxedo} demonstrated that PoW miners who bypass validation stand a significantly higher chance of winning the block compared to their mining power.
Despite the similarity in concept, our work focuses on \textit{transaction} validations in the P2P network.
In our scenario, the centralized entities (i.e., validators, relays) forward transactions without validations, giving users extra time to execute transaction re-ordering attacks~\cite{daian2020flash, qin2022quantifying}. 
As long as there is an economic incentive to capture transactions fast, centralized entities strive to minimize latency, which 
might make the P2P network more centralized and introduce new security vulnerabilities.

Our proposed attack stems from not only DoS attack~\cite{kumar2007smurf} but also Economic Denial of Sustainability (EDoS) attack. 
Hoff~\cite{hoff2008edos} first presented the EDoS concept in a 2008 blog post.
EDoS, akin to a traditional Denial of Service (DoS) attack, focuses on inflicting financial losses by targeting the victim's traffic usage~\cite{wang2016abusing}.
Attackers exploit the knowledge of a victim server hosted on a cloud service by manipulating traffic usage to inflate bills, making the victim's operation economically unsustainable.
The same issue applies to blockchain P2P networks where many nodes have public IP addresses, and use cloud platforms such as AWS.

\section{Data}
\label{sec:data}
We use publicly available data (RPC nodes, txpool providers) from Flashbots~\cite{flashbot_md_data}, as well as measurement through our nodes deployed in a P2P network (Figure~\ref{fig:spynode}). 

\subsection{Public txpool data}
\label{subsec:public_data}
Each node has a different view of the txpool, meaning that they hold a different set of pending transactions. 
A node that opens up its API connection to the public makes its txpool accessible  (\texttt{newPendingTransactions}).
Flashbots\footnote{\url{https://www.flashbots.net/}} has been providing publicly available txpool information on ``Mempool Dumpster''~\cite{flashbot_md_data} every day, and we use its data from September 1st, 2023 to January 11th, 2024.  
Flashbots collects a set of pending transactions from 1) RPC providers and generic nodes -- Infura, Alchemy, A-pool,\footnote{A regular network node with the optimized peering setting whom Flashbots listens to.} Flashbots' local node, Mempoolguru~\cite{mempoolguru}, and 2) the infrastructure txpool providers -- bloXroute, Chainbound, Eden. 
All the transactions contain transaction hash, source (i.e., which service/node), and observation timestamp (in milliseconds). 
Some portions of transactions include the details of transactions (e.g., sender, receiver, gas, value, and data fields). 
We remove repetitions (identical transactions) over time to avoid double-counting.  
Consequently, our summary statistics slightly diverge from Flashbots'.

\subsection{Private monitoring node data}
\label{subsec:private_data}
\begin{figure}[tb]
    \centering
    \includegraphics[width=0.8\linewidth]{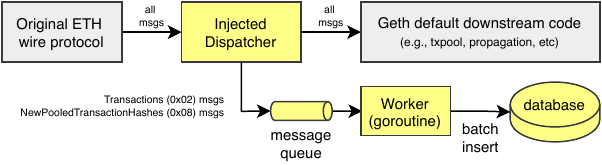}
    \caption{An overview of our monitoring node.}
    \label{fig:spynode}
\end{figure}

To infer the network topology and estimate the attack impact, we modify Geth to store all messages received on the P2P network layer between September 1st, 2023 and January 25th, 2024. 
We deploy two nodes in the network to 1) avoid a single point of failure and 2) increase the coverage and robustness of our estimates.  
Those monitoring nodes enable us to observe transactions before they enter the txpool as described in Figure~\ref{fig:spynode}; 
modified components are highlighted in yellow, whereas the standard Geth components are in grey. 
We have integrated a custom message dispatcher that filters and timestamps specific transaction propagation messages (e.g., \texttt{0x02} and \texttt{0x08} messages). 
These messages are subsequently forwarded to a message queue, then batch-processed by a goroutine worker for database entry. 
This configuration ensures there is no delay between the capture of a message and its logging in the database.
For each transaction hash, we only record the first message to reduce the size of the dataset. 
We store the type of messages, the content of the transactions, the origin node IDs, and the timestamp.  
This is critical for 1) estimating the active peer connections (\S\ref{subsec:estimate_gx}), and 2) identifying spamming behavior and modified nodes within the P2P network (\S\ref{subsec:modified_nodes}).  
Our two Geth nodes operate on Ubuntu 20.04.2 LTS systems, one with an AMD Ryzen Threadripper 3990X (64-core, 2.9 GHz) and the other with a 13th Gen Intel(R) Core(TM) i9-13900KF. 
Both machines have 256~GB and 64~GB of RAM, respectively, and are supported by NVMe SSDs.  
We raise the limit of peer connections in both Geth clients to 1,000. 
Both nodes are located in Europe and are operated independently.
We keep the peering default configuration (no change in whitelist/blacklist).
In total, we capture about 2.5 billion (2,493,695,017) unique \texttt{0x02} and \texttt{0x08} messages from 36,815 peers in the aforementioned timeframe.

To further profile each node within a network, we gather peer information from our nodes. This is done by subscribing to the \texttt{admin\_peerEvents} API,\footnote{\url{https://geth.ethereum.org/docs/interacting-with-geth/rpc/ns-admin}} which provides details such as node names, software, network IDs, compatible protocols, and IP addresses. 
This API alerts peer events to the subscriber in real-time, capturing both the addition and removal of peers, including those that connect/disconnect briefly.
This information helps us calculate the number of active peer connections at each time, and better infer peer stability. 
For IP addresses, we call the \url{ipinfo.io} API\footnote{\url{https://ipinfo.io/developers/responses}} to obtain additional information such as Internet Service Provider (ISP) or Autonomous System (AS), hostname, timezone, region, city/country, and registered location. 
Besides regular full nodes, we also use our archive node to access the full state of the blockchain. 
In particular, we 1) retrieve the account balance/nonce at specific times
(\S\ref{subsec:characterize_empirical_spam}) and 2) fork the archive node to simulate
the transaction validation process (\S\ref{subsec:val_simulation}).
To accurately assess the attack cost amid price fluctuations, we
obtain historical Ethereum/USD exchange rates through the Coingecko
API.\footnote{\url{https://www.coingecko.com/api/documentation}}

\section{Modeling amplification}
\label{sec:model_edos}
We turn to formally modeling the Blockchain Amplification Attack in detail. 

\subsection{Threat model}
\label{subsec:model_setting}
We first formalize an amplification attack on the blockchain
P2P network (Figure~\ref{fig:edos_sketch}). The attack
leads modified nodes (i.e., victims) to forward invalid transactions,
and increases their outgoing traffic usage and cost. The attacker attempts 1) to degrade the quality of the service provided by modified nodes, 
2) to cause economic damage to modified
nodes, and 3) to disrupt all nodes and users (especially MEV bots or block builders that listen to the services provided by modified nodes). 
Beyond blockchain
ecosystem participants, cloud service providers also potentially have
an incentive to support attacks on modified nodes to increase traffic
usage and, consequently, revenue.

\begin{figure}[]
    \centering
    \includegraphics[width=0.8\linewidth]{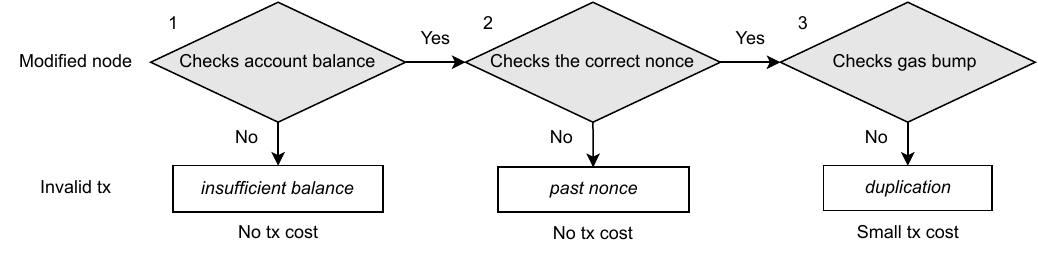}
    \caption{Invalid transaction creation}
    \label{fig:invalid_tx}
\end{figure}

We next delineate the attack process.
The attacker follows the procedure in Figure~\ref{fig:invalid_tx} to produce an invalid transaction, $tx_{invalid}$, of size $a$~bytes. 

\begin{enumerate}
    \item If the target modified node does not check the account balance, the attacker can generate new addresses and send an infinite number of invalid transactions. 
    The transaction cost is zero since there is no chance of inclusion.
    \item If the modified node checks the balance but not the nonce, the attacker can send transactions with a previously used nonce, which also costs nothing. One necessary condition is for the attacker to rely on an account with a long transaction history (i.e., having a high nonce allows the attacker to reuse many past nonces). 
    \item If the modified node checks balance/nonce but does not use the gas bump rule (\S\ref{sec:bg}), the attacker can replicate transactions while maintaining the same sender, nonce, and gas price. The attacker slightly alters one parameter (e.g., a data field) and re-signs the transaction to yield a different hash.
    The attacker could incur transaction costs if the original (or one of the duplicated ones) ends up on the chain.
\end{enumerate}

In the first two cases, the attacker can specify an unreasonably high gas price to prioritize their transactions in the txpool, without additional costs (because those transactions will be invalidated later). 
The attacker can also combine multiple strategies (e.g., duplicating many invalid transactions using a past nonce). 

Next, the attacker selects its target modified node.
If the attacker receives any invalid transactions from a given peer, that peer is most likely the modified node.
Alternatively, the attacker can monitor various centralized entities' txpools, identify the service(s) that accept(s) invalid transactions, and send $tx_{invalid}$ through RPC endpoints. 
The attacker finally sends $tx_{invalid}$ to one modified node, which accepts/inserts $tx_{invalid}$ to its txpool and forwards it with the rest of the network, including other modified nodes and regular nodes. 

The modified nodes are tyipcally run by the same entity (as empirically observed in \S\ref{subsec:modified_nodes}), they are most likely connected to minimize latencies.
Hence, $tx_{invalid}$ would reach all the modified nodes. 
Even if some modified nodes belong to multiple entities, $tx_{invalid}$ would reach out to the rest as long as one of the nodes from each entity connects.

Regular nodes keep receiving $tx_{invalid}$ as a \textit{new} transaction because $tx_{invalid}$ never gets into their txpool.
All nodes consume an incoming traffic and CPU resources for transaction verification, illustrating the attack severity.  

\subsection{Amplification factors}
\label{subsec:model_amplifiation}
We want to precisely estimate the \textit{amplification factor} to show how the attack scales up given the attacker's input. 
The DoS/EDoS literature uses two metrics to quantify attack effectiveness: Traffic Amplification Factor (TAF)~\cite{kumar2007smurf} and Economic Amplification Factor (EAF)~\cite{wang2016abusing}, 
defined as 
$$
TAF = \frac{B_{victim}}{B_{attacker}}, \quad EAF = \frac{\lambda_{victim}}{\lambda_{attacker}} \ ,
$$
where $B_{victim}$, $B_{attacker}$ is the traffic from the victim and attacker node, respectively.
In our scenario, TAF is the ratio of outgoing traffic generated by the attacker to the modified nodes, which corresponds to the original red arrow and the sum of the blue arrows in Figure~\ref{fig:edos_sketch}, respectively. 
We calculate TAF for the network but not for a single modified node.
$\lambda_{victim}$ (resp. $\lambda_{attacker}$) is the outbound traffic cost that the victim (resp. attacker) node pays for providers: in our case, all the modified nodes (resp. a single attacker). 
We can also re-write EAF as
\begin{equation}
\label{eq:eaf}
EAF = \frac{B_{victim} \cdot p_{victim}}{B_{attacker} \cdot p_{attacker}} = TAF \cdot 
\frac{p_{victim}}{p_{attacker}}   \ ,
\end{equation}
where $p$ is the price per outgoing traffic cost based on the ``pay-as-you-go'' policy of cloud services.
Typically, it should be easier for the attacker than for the victim to minimize their costs. 
Given that the attacker knows the IP address (i.e., location) of the modified nodes, they can launch the attack in the same data center and reduce their cost. 
This discrepancy (``price multiplier'') amplifies EAF. 

\subsection{Network waste}
\label{subsec:network_waste}
To estimate how the original attack traffic amplifies in the network, we first derive 1) the number of bytes a \textit{regular} node $i$ receives based on the number of peer connections (see blue arrows directed toward \textit{each} regular node in Figure~\ref{fig:edos_sketch}), 2) the number of bytes the entire network wastes based on the distributions of active connections on each peer (all the blue arrows). 
This calculation only considers the public nodes in a P2P network, but not the users (MEV bots and block miners/builders) who 1) privately connect to txpool providers or 2) listen to transactions through RPC endpoints outside the P2P network. 
The actual amplification factor would be the combination of 1) the amount of traffic that goes to users who listen to modified nodes and 2) TAF.  
We indeed notice that some modified nodes shut down the outbound connections to reduce outbound costs. Yet, they still incur the cost of propagating invalid transactions to users.

A regular node only hears $tx_{invalid}$ from modified nodes (not from other regular nodes). 
This allows us to focus solely on the connections from modified nodes to each regular node.
Let the number of active connections for each regular node $i$ be $x_i$. 
The number of the modified node connection is $x_i \gamma$ where $\gamma$ is the ratio of modified nodes.
This is based on the assumption that the node discovery process is random, thus the number of connected modified nodes increases linearly along with the number of connections by expectation. 
As described in \S\ref{sec:bg}, Ethereum nodes broadcast the transactions in two ways: 1) \texttt{0x02} message called ``broadcast'' ($a$ bytes), and 2) \texttt{0x08} message called ``announcement'' ($32$ bytes). 
Although Ethereum Wire protocol~\cite{ethereum_wire} suggests propagating \texttt{0x02} to a small number of peers for network efficiency, there is no clear consensus on the implementation, leading to variations across different software clients. 

For Geth, the propagation policy is consistent across versions; the node propagates the broadcast message (\texttt{0x02}) to a square root of connected nodes ($\sqrt{x_i}$) and just sends the announcement message (\texttt{0x08}) to the rest ($x_i - \sqrt{x_i}$)) defined in the function \href{https://github.com/ethereum/go-ethereum/blob/master/eth/handler.go\#L627}{\texttt{BroadcastTransactions}}. 
We call this the ``\emph{square root policy}.''
The node picks up the subset of nodes uniformly randomly for \textit{every} transaction.

In Erigon, just like Geth, the node broadcasts (\texttt{0x02}) a square root of connected nodes, and sends an announcement (\texttt{0x08}) to \textit{all} peers (not the rest of the peers).
While Erigon adopted a different propagation strategy after v2.49.0, we only focus on the square root policy given the small number of nodes after v2.49.0.
The description of the change in Erigon propagation strategies is in Appendix~\ref{appendix:erigon_prop_policy}.

\begin{figure}
  \begin{subfigure}{0.4\linewidth}
   \includegraphics[width=\textwidth]{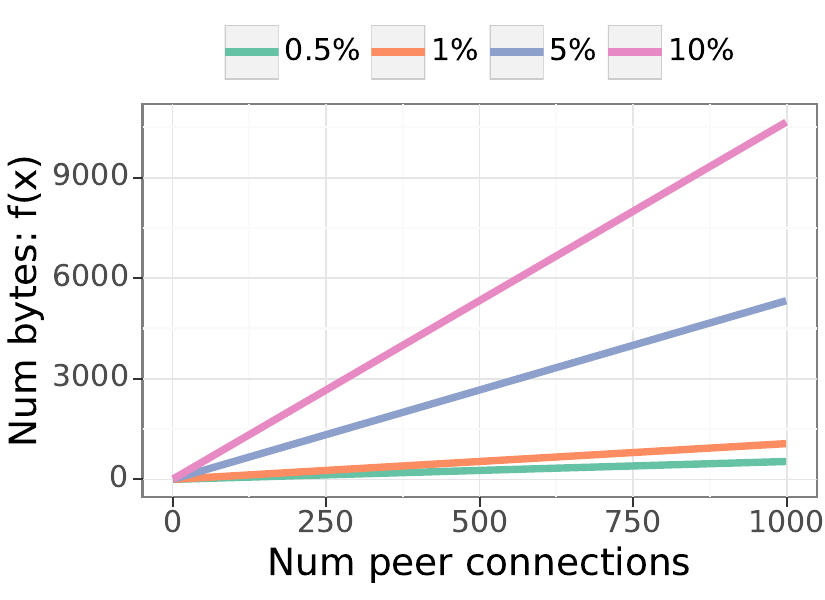}
   \caption{modified node ratio}
   \label{subfig:amp_adv_ratio}
  \end{subfigure}
  \begin{subfigure}{0.4\linewidth}
    \includegraphics[width=\textwidth]{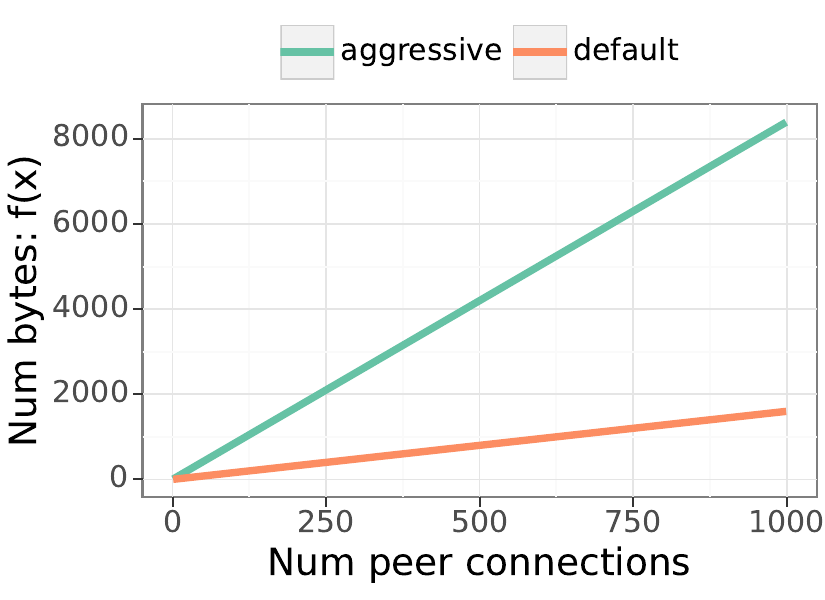}
    \caption{Propagation policy}
    \label{subfig:amp_prop_type} 
  \end{subfigure}
  \caption{Number of bytes a regular node $i$ receives based on the number of active connections.}
  \label{fig:node_i_waste} 
\end{figure}

We here simply assume that the modified nodes hold 50~connections and send \texttt{0x02} messages to 14\% of them ($\sqrt{50}/50=0.141$), and send a hash to the rest.
By expectation, the node receives $0.14 a + 0.86\times 32 = 0.14 a + 27.52$ bytes for one modified node connection. 
However, we also empirically observe that some nodes neglect the square root policy and broadcast only \texttt{0x02} messages ($a$ bytes) to every peer (no \texttt{0x08} message).
In this scenario, the node consistently receives complete transactions instead of just hashes, eliminating the need for subsequent communication such as transaction content requests (\texttt{0x09}) from peers. 
This could also potentially result in latency reduction.
We denote those two types of propagation policies as $\pi=\text{``sqrt'' and ``aggressive''}$, respectively.
We multiply by the number of modified nodes ($\gamma x_i$) in the connection, so the total amount of waste regular node $i$ receives is 
$$
f(x_i)= 
\begin{cases}
  (0.14 a + 27.52) \gamma  x_i  & \pi = \text{``sqrt''} \ , \\
    a \gamma x_i & \pi =\text{``aggressive''} . 
\end{cases}
$$
Figure~\ref{fig:node_i_waste} illustrates the change in $f(x)$ based on the number of peer connections.
The left figure alternates the ratio of modified nodes ($\gamma$), and the right one shows two types of propagation ($\pi$). (Default: $a=560$, $\gamma=0.015$, $\pi=\text{``sqrt''}$.) 
The change in $\gamma$ or $\pi$ affects the amount of waste linearly. 

We next calculate the total network waste by considering the distribution of peer connections, which we empirically estimate in \S\ref{subsec:estimate_gx}.
There are $(1-\gamma) \cdot N = N_{\mathcal{R}}$ regular nodes where $N$ is the total number of nodes in the network. 
We accumulate the amount of waste for each node $f(x)$ to get the overall network waste: $\sum_{i=1}^{N_{\mathcal{R}}} f(x_i)$. 
In our measurement, we estimate $x$ for a part of the network (i.e., our connected peers), and apply smoothing to derive the continuous distribution $g(x)$. 
In that case, the network waste is $N_{\mathcal{R}} \cdot \int_{0}^{1000} f(x)g(x)dx$, where the second term is the expected waste per node. 
Here, we assume a maximum connections of 1,000 to ensure convergence and exclude an impractically large connections. 
We finally divide this by the transaction size ($a$) to get the amplification factor, TAF. 
An adversary could set up modified nodes on their own to increase $\gamma$, but these nodes would also be attacked, making this strategy unproductive.

\section{Empirical examples of attack}
\label{sec:empirical_spam}
This section identifies the similar attack instances seen in the wild, and 
characterizes the size, intensity, cost, and strategies, and actors
(the attacker -- sender address, the victim -- centralized entities)
involved in those attacks. 
While we cannot definitively pinpoint the motivations of those attackers, those attacks are executed in
a manner consistent with our proposed attack (see the discussion in
Appendix~\ref{appendix:intention}).

\subsection{Victims}
\label{subsec:compare_centralized}
\begin{figure}[]
  \begin{subfigure}{0.45\linewidth}
   \includegraphics[width=\textwidth]{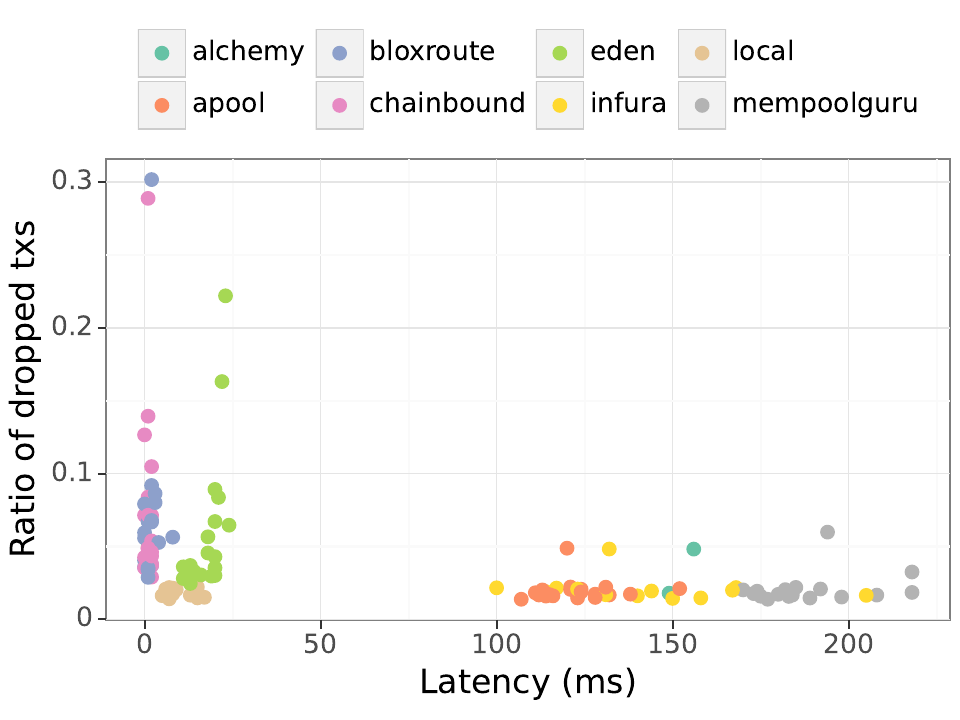}
  \end{subfigure}
  \begin{subfigure}{0.4\linewidth}
   \includegraphics[width=\textwidth]{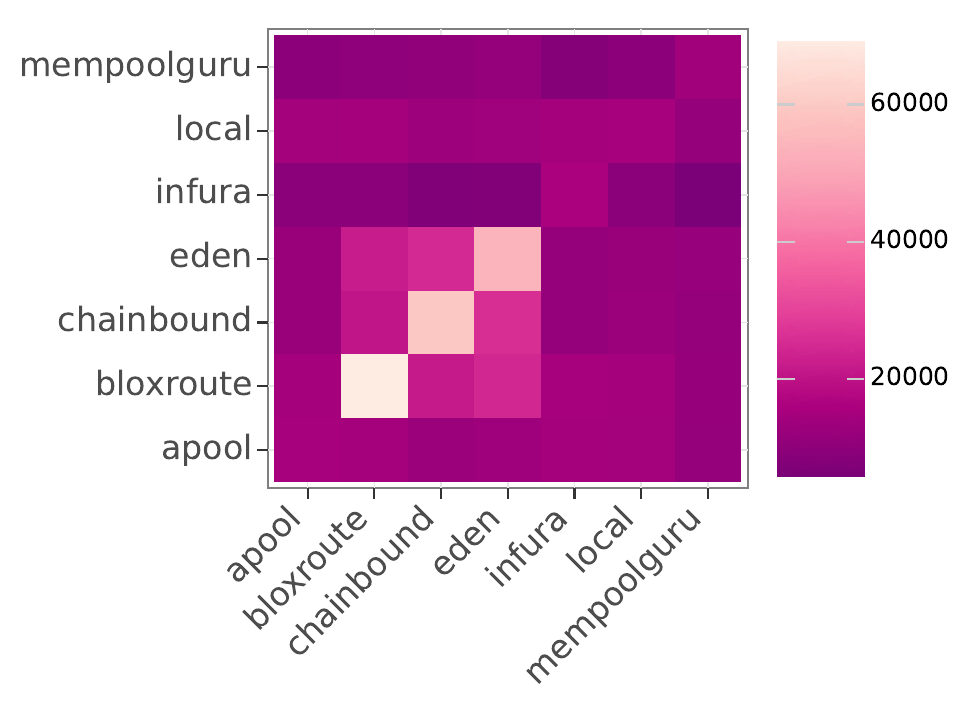}
  \end{subfigure}
  \caption{Comparing the centralized entities' txpools (left: accuracy-latency plot, right: dropped transactions between sources.}
  \label{fig:centralized_services}
\end{figure}

We first attempt to find the instances of the attack in the txpool of centralized entities as they are often the target of the attack. 
We find that some services deliver transactions much faster than others, but also propagate a significant number of dropped transactions.
By comparing their txpools (from Flashbots' data), we can practically figure out how fast, and how accurate their services are. 
For the latency comparison, we use the timestamp recorded in Flashbots' data. 
For each on-chain (finalized) transaction, we check how late each source propagates from the first moment Flashbots sees the transactions from any source.
For accuracy, we calculate the ratio of the number of dropped transactions to the number of total transactions received. 
These numbers convey the quality of their propagation flow. 
Figure~\ref{fig:centralized_services} (left) illustrates the bi-modal representation of latency vs. accuracy except for Flashbots' ``local'' node. 
Each data point represents one week of data for each service.  
The $x$-axis is the median latency from all the transactions, whereas the $y$-axis is the ratio of dropped transactions.
Services such as bloXroute, Chainbound, and Eden invest in infrastructure to optimize their propagation flow, resulting in potential latency reduction.
Moreover, Flashbots' database connects to bloXroute and Eden through gRPC, which is faster than the web socket commonly used by others (``delivery latency''). 
Additionally, a geographical advantage exists for services located close to the database---e.g., Flashbots' local node likely operates a physical proxy to its database.
The figure also suggests that those ``fast'' services fail to filter out transactions that are not supposed to persist in the network. 

Figure~\ref{fig:centralized_services} (right) is a co-occurrence matrix where each cell indicates the number of dropped transactions common to both services (the $x$- and $y$-axis) normalized by each day. 
The diagonal (from the bottom left to the top right) is the number of dropped transactions to each source---e.g., bloXroute has around 60,000 new dropped transactions per day.  
bloXroute, Chainbound, and Eden collectively accept a significantly high number of dropped transactions that are not seen in any of the other txpools, which suggests that they may have employed different propagation strategies from others. 
Those results hypothesize that a part of the latency reduction might also come from transaction validation process, which motivates our simulations in \S\ref{subsec:val_simulation}.

\subsection{Identifying and characterizing attack behavior}
\label{subsec:characterize_empirical_spam}
We find that some Ethereum accounts send an unexpectedly large number of transactions in a short range of time. 
We set two (conservative) thresholds to extrapolate automated non-human spamming behavior (as a lower bound): if 1) we detect over 100 transactions from a single sender within a single block interval (12 seconds) and 2) more than 95\% of those transactions fail to be included in the chain.
The second check is necessary to filter out a large entity (e.g., a popular gambling site) that submits many transactions at once without spamming intent. 
Based on this heuristic, we find 2,591 instances from 345 Ethereum addresses from Flashbots data from September 1st, 2023 to January 11th, 2024.

We next characterize attack like behavior, in particular, three types of invalid (attack) transactions discussed
in \S\ref{subsec:model_setting} and Figure~\ref{fig:invalid_tx}: 1) insufficient balance, 2) past
nonce, and 3) duplicate transaction. 
To check the account balance and nonce
at the time of transaction submission, we
use our archive node to fetch the state of the blockchain at the
time of the timestamp. 
Given a network delay between
when the transaction enters a P2P network and when Flashbots records
the timestamp, 
we look at the state of the account
two blocks before the observation timestamp (i.e., \textit{at least} 12 seconds in the past). 
Of 2,591 attack cases,
we confirm that 536 instances (from 224 unique addresses) come from
accounts that do not have a sufficient balance.
Two services (bloXroute and Eden) have included those
transactions during our observation window. In particular, some
addresses do not have any transactions (send or receive) in their
history whatsoever (i.e., zero Ether balance).
In addition, three services (Chainbound, Eden, bloXroute) have accepted transactions that
are using a past nonce (and thus, are invalid) during our measurement
interval. Just like we identify invalid transactions from
accounts with insufficient balance, we retrieve the last nonce of the corresponding
address and identify 62 attacks from 27 unique addresses that use previously used nonce.

Attackers also duplicate transactions in order to generate many
invalid transactions. We categorize our identified attack instances
based on the transaction parameters the attackers manipulate the
most. There are four main parameters the attacker calibrates; 1)
the gas limit, 2) the nonce, 3) the data field, and 4) the transfer value. This does not necessarily mean that the attackers only change
one specific parameter, but sometimes they simultaneously permute
several sets of parameters to increase the invalid transaction set,
including other variables such as the recipient's addresses, chain
ID, and access list (introduced in EIP-2930).
Table~\ref{tab:dos_type} summarizes the key statistics such as the
transaction costs, intensity, and the size of the attack for each
category. We first define key metrics: the size, cost, and intensity of the attack.

\textbf{\textit{Size}}: we count the average number of transactions we
observe during the attack and calculate the average transaction size
(an RLP-encoded raw transaction in bytes). Since we only consider the
transactions that enter one of the txpools Flashbots listens to, the
number of transactions generated/sent by the attacker should be strictly
larger.

\textbf{\textit{Transaction costs}}: we calculate the ``cost'' of the
attack as the amount of transaction fee (i.e., the effective gas price
multiplied by the amount of gas used) for the original (on-chain)
transactions that the attacker duplicates. For better comparison, we
fetch the price data of Ethereum at the time of the attack and convert
the total amount of Ether to USD.

\textbf{\textit{Intensity}}: we compute the attack intensity based on the median of the timestamp interval (i.e., the time lag
between two consecutive transactions observed in the txpool).

About 63\% of attack instances calibrate the \textit{gas limit} parameter, but
only two addresses are involved. Those two addresses manage to send
1,581 transactions with only 0.26 milliseconds between transactions on
average. The largest attack carries 23MB of \textit{unique} data between blocks (12 seconds). 
Since, on average, only one
transaction lands on the chain, the transaction cost to the attacker is around 3~US
dollars.
Chainbound seems to be the main victim
of this attack.

Changing nonce (future/past) is also a common practice for attackers
to generate a large number of invalid transactions. 300 addresses do
so over 784 cases. Each of these attacks carries 227 transactions
on average. Given that almost none of these transactions end up
being included in the chain, the cost is exceedingly small. 

The data field/value appears to be more costly than other methods (more
than 30 USD) and carries a smaller number of transactions with around 10
milliseconds of interval. However, more (2.71 on average) services constantly accept those
invalid transactions.

To sum up, Table~\ref{tab:dos_type} illustrates each strategy's pros
and cons; the attacker can manipulate different parameters to prepare many invalid transactions,
depending on what the attacker wants to maximize/minimize: costs, size,
intensity, and the number of victims (i.e., the centralized entities).

\begin{table*}[]
\centering
\footnotesize
\begin{tabular}{@{}lllllllll@{}}
\toprule
Type  & \# of cases & addr & \begin{tabular}[c]{@{}l@{}}\# of txs \\ (avg.)\end{tabular} & \begin{tabular}[c]{@{}l@{}}\# of on-chain tx\\ (avg.)\end{tabular} & \begin{tabular}[c]{@{}l@{}}Total cost\\ (avg. in USD)\end{tabular} & \begin{tabular}[c]{@{}l@{}}Txs interval\\ (avg. in ms)\end{tabular} & \begin{tabular}[c]{@{}l@{}}Txs size\\ (avg. in bytes)\end{tabular} & \begin{tabular}[c]{@{}l@{}}\# of victims \\ (avg.)\end{tabular} \\ \midrule
gas   & 1719        & 2    & 1581.00                                                     & 0.92                                                               & 3.26                                                               & 0.26                                                                & 1496.2                                                             & 1.04                                                            \\
nonce & 784         & 300  & 227.24                                                      & 0.03                                                               & 0.05                                                               & 9.22                                                                & 120.2                                                              & 1.09                                                            \\
data  & 71          & 30   & 167.23                                                      & 1.11                                                               & 31.97                                                              & 10.56                                                               & 563.0                                                              & 1.70                                                            \\
value & 17          & 13   & 116.35                                                      & 0.94                                                               & 132.00                                                             & 7.91                                                                & 481.7                                                              & 2.71                                                            \\ \bottomrule
\end{tabular}
\caption{Summary statistics for each attack type.}
\label{tab:dos_type}
\end{table*}

\section{Estimating model parameters}
\label{sec:estimate_param}
We next derive our model parameters, which we then use to calculate amplification factors. 

\subsection{Peer connection distribution: $g(x)$}
\label{subsec:estimate_gx}
We formalize our method for estimating the number of active connections
for each peer, $x_i$, and empirically computing its distribution,
$g(x)$. Although studies on blockchain P2P network topology exist,
the key novelty in our approach is to estimate \textit{active} peer connections
on the Ethereum \textit{mainnet} without submitting test transactions (i.e.,
mediating ethical concerns and research costs). 
Instead, we reverse-engineer the number of peers based on propagation message types.
We collect 2.5 billion messages in five months (as explained in \S\ref{subsec:private_data}) to perform this method.

As outlined in \S\ref{sec:model_edos}, an unmodified Geth node $i$ has a number of peer connections $x_i$ and \textit{uniformly randomly} chooses $\sqrt{x_i}$ nodes to broadcast (\texttt{0x02}) transactions, 
while sending announces (\texttt{0x08}) to the rest. 
Our node receives a \texttt{0x02} message with probability $\theta_i = \frac{\sqrt{x_i}}{x_i}$, which means the data distribution follows a \textit{binomial} distribution with parameters $\theta_i$ and $m$, where $m$ is the number of messages we receive from each peer.
Specifically, $m = m_2 + m_8$ where $m_2$ and  $m_8$ are the number of \texttt{0x02} and \texttt{0x08} messages sent by peer~$i$, respectively. 
The probability (likelihood function) of the binomial distribution is defined as 
\begin{equation}
\label{eq:likelihood}
l(\theta_i, m_2, m_2 + m_8) = {m_2 + m_8 \choose m_2} \theta_i^{m_2} (1-\theta_i)^{m_8} \ .
\end{equation}

We derive an unbiased maximum likelihood estimate $\hat \theta_i$ by taking the derivative of the log-likelihood function, and reconstruct the number of peers $x_i$ as: 
\begin{equation}
\label{eq:sqrt_policy_geth}
\hat \theta_i= \frac{m_2}{m_2 + m_8} \Leftrightarrow \hat x_i = \left(\frac{1}{\hat \theta_i}\right)^2 = \left(\frac{m_2 + m_8}{m_2}\right)^2 \ .
\end{equation}

Since the estimate $\hat x_i$ becomes unreliable with smaller samples $m$, we also derive the variance of $\hat \theta$ and use it to exclude unreliable estimates on $x_i$. 
This is likely to exclude nodes with many peer connections, which require a larger sample to reduce the variance; thus we might underestimate the expected value of $g(x)$.
However, 1) the percentage of nodes with large connections is disproportionately small, 2) the exclusion process does not lead to overestimating amplification factors, and 3) our large dataset helps alleviate this shortcoming. 
The details of how we derive the equations above and exclude unreliable estimates are in Appendix~\ref{appendix:mle_estimate}.

\begin{figure}
  \begin{subfigure}{0.4\linewidth}
   \includegraphics[width=0.9\textwidth]{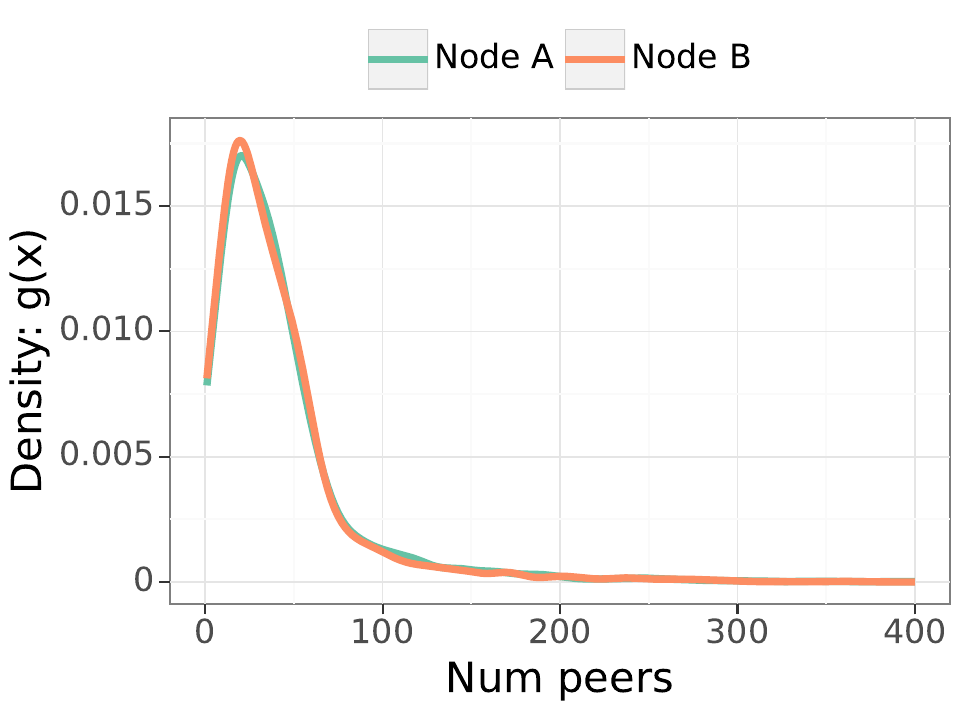}
   \label{subfig:pdf_peer_connection}
  \end{subfigure}
  \begin{subfigure}{0.4\linewidth}
    \includegraphics[width=0.9\textwidth]{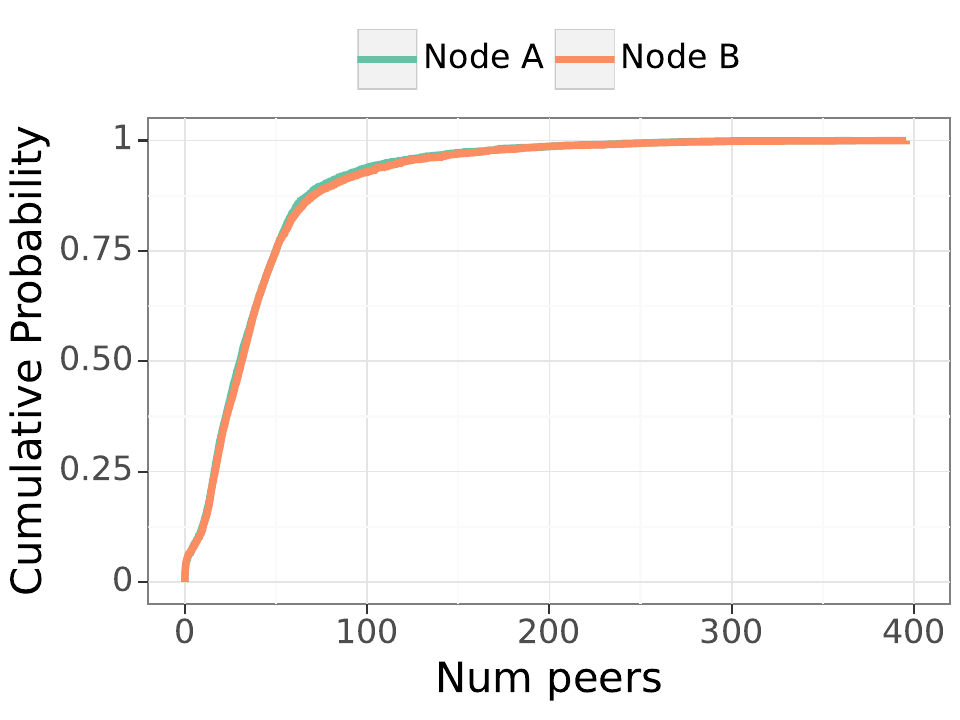}
    \label{subfig:cdf_peer_connection} 
  \end{subfigure}
  \caption{Number of active peer connections for each regular node: left (PDF with a KDE smoothing), right (CDF).}
  \label{fig:gx_peer_connection} 
\end{figure}

We apply Eqn.~(\ref{eq:sqrt_policy_geth}) for peers (of \textit{our} nodes) that 1) use Geth or Erigon (before v2.49.0), and 2) we exclude nodes with a non-public \texttt{git commit} string, which is a sign of software  as we discuss in \S\ref{subsec:modified_nodes}. 
If our two monitoring nodes observe the same peer, we take the average between two. 
In total, we manage to estimate the number of peer connections for 6,005 nodes: mean and median of 41 and 31, respectively. 
For those, each peer has sent an average (median) of 293,121 (81,327) transaction messages.
To make the distribution generalized/smooth and get $g(x)$, we apply the kernel density function (KDE) on SciPy's implementation \& Scotts' method for the bandwidth selection.
Figure~\ref{fig:gx_peer_connection} shows the resulting PDF (Probability Density Function) and CDF (Cumulative Density Function) of active peer connections (green: estimate from Node A, orange: from Node B).
The strong alignment between the two estimates indicates our method is robust. 
As an additional check, we ask one Ethereum node operator to provide us with the node name, node ID, and the maximum number of peers. 
The node uses an unmodified Geth node and sets the maximum number of peers to 40.
Our estimator tells us that this node is expected to have $33.68$ peers.
Given that not all the nodes are active/stable over time, our estimate appears 
to be consistent with this piece of ground truth.

\subsection{Ratio of modified nodes: $\gamma$}
\label{subsec:modified_nodes}
We estimate the ratio of the modified nodes ($\gamma$) from the set of peers we connect.
As described in \S\ref{subsec:private_data}, we capture transactions before entering the txpool, so we can empirically label the modified nodes that forward invalid transactions to us. 
First, we prepare three sets of invalid transactions as introduced in \S\ref{subsec:model_setting} and \S\ref{sec:empirical_spam} to examine peers' validation process: 1) insufficient balance, 2) past nonce, and 3) duplication (gas bump rule). 
We first label peers that have ever sent invalid transactions of the first two kinds: insufficient balance and past nonce.
Our approach may mislabel modified nodes if their blockchain status is unsynced. 
However, the fact that the node forwards invalid transactions still allows attackers to conduct the same attack on unsynchronized nodes. 

With regard to duplications, we prepare the set of duplicated transactions (e.g., changing the data field while keeping other parameters constant.) 
We then label peers that forward two duplicated transactions
within one second, which indicates that the duplicated transactions stay
in each peer's txpool at the same time. That fact suggests the absence
of a gas bump rule (or that the threshold was set to zero).

Surprisingly to us, we do not find many nodes that propagate
transactions without sufficient balances or with past nonces even though
we observe many at the RPC endpoints. We conjecture that those nodes
might have already adopted a technique similar to the second mitigation
we propose in \S\ref{subsec:limitation} or restricted the outbound
traffic.

To derive the ratio of modified nodes in the duplication case, we look at the set of nodes in our active connections and check whether any of those are from modified nodes. 
Figure~\ref{fig:num_vul_peers_ratio} shows the ratio of connections to modified nodes at each snapshot (hourly) between December 15, 2023, and January 10, 2024.
Both nodes $A$ and $B$ 
maintain approximately 15 connections over time with
modified nodes; dividing the number of total number of connections
(roughly 1,000) at each time interval gives us an average ratio of about
1.5\%. 
In total, we identify 175 unique modified nodes.

Based on the HEAD of the \texttt{git commit} in the node name (which is generated when compiling the source code), we determine whether the node uses the publicly available version of the software or the private forks of the repository.
174 out of 175 peers share an identical git commit hash, which is not present in the corresponding GitHub repository. (We fail to collect data for one remaining node.) 
This fact suggests that those nodes are likely from the same entity and ensures that an invalid transaction reaches out to all modified nodes. 
The details of how we extract the \texttt{git commit} hash and use it to
determine whether a node was customized 
can be found in Appendix~\ref{subsec:customized_nodes}. 
Further, we find that these modified nodes 
apply an aggressive propagation strategy ($\pi$=``aggressive'') -- i.e., 
they only broadcast (\texttt{0x02}), but never announce (\texttt{0x08}), further increasing amplification effects.

\begin{minipage}{.97\linewidth}
  \begin{minipage}[]{0.35\linewidth}
      \centering
      \includegraphics[width=\linewidth]{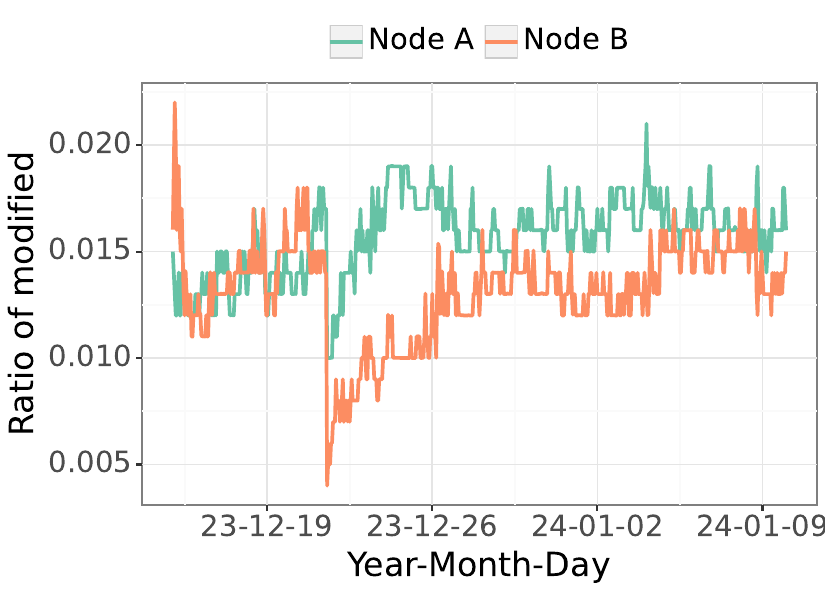}
      \captionof{figure}{Ratio of modified peers in our nodes' connections.}
      \label{fig:num_vul_peers_ratio}
  \end{minipage}
  \hfill
  \begin{minipage}[]{0.58\linewidth}
    \centering
    \scalebox{0.75}{
    \begin{tabular}{@{}llll@{}}
    \toprule
    Parameter     & Description                           & Estimate     & Source \\ \midrule
    a             & Tx size                               & 560          & \S\ref{subsec:characterize_empirical_spam} \\
    $\gamma$      & Ratio of $\mathcal{M}$ modified nodes & 0.015        & \S\ref{subsec:modified_nodes}       \\
    $\pi$              & Propagation policy                    & ``aggressive'' & \S\ref{subsec:modified_nodes}       \\
    N             & Number of nodes                       & 6000         & \cite{etherscan, ethernodes}       \\ \midrule
    Per tx        & per regular / modified node           & 345 / 22,640 &        \\
                  & Entire network                        & 2,037,613    &        \\ \midrule
    Amplification & BAF                                   & 3638         &        \\
                  & AWS ratio                             & 0.2          & Appendix~\ref{appendix:software_isp}       \\
                  & EAF                                   & 13827        &        \\ \bottomrule
    \end{tabular}
    }
    \captionof{table}{Estimated parameters for the model.}
    \label{tab:model_param}
  \end{minipage}
  \end{minipage}

\subsection{Amplification factor}
\label{subsec:estimate_amplification}
Based on the estimated parameters in the above sections, we finally
calculate the amplification factors. 
We derive TAF based on the
transaction duplication case, but the same technique could apply to insufficient balance or past nonce cases. Table~\ref{tab:model_param} summarizes the
model parameters and the final results. From the top, we refer to the
size of transaction $a=560$ (bytes) in \S\ref{subsec:characterize_empirical_spam},
modified node ratio $\gamma=0.015$,
and propagation policy $\pi$=``aggresssive'' in \S\ref{subsec:modified_nodes}. We set the number of
peers $N$ to 6,000 based on the statistics reported by the third-party websites \cite{etherscan,
ethernodes} after excluding non-Ethereum nodes. 
All in all, we estimate the total waste per transaction in the
network to be 2,037,613 bytes (2.0 MB). 
Given that the transaction size
($a$) is 560, the amplification factor is 3,638.

To quantify the economic impact of network waste (EAF), we next calculate the financial loss of spreading invalid transactions at modified nodes.
We observe that many centralized entities deploy their nodes in the cloud service (Appendix \ref{appendix:software_isp}), so we calculate the data transfer cost based on the ``pay-as-you-go'' policy of the cloud services.
We choose AWS pricing\footnote{\url{https://aws.amazon.com/ec2/pricing/on-demand/\#Data\_Transfer\_within\_the\_same\_AWS_Region}} as an example.
AWS does not charge for inbound traffic but for outbound traffic (plus, significantly less for the traffic within AWS).
Most cloud services follow a similar standard.
If we assume that the cost of inbound traffic is zero, the attack incurs costs only for 1) the attacker (to send invalid transactions) and 2) the modified nodes (to propagate invalid transactions to the rest).

We compute EAF (economic amplification factor) based on TAF and the traffic price ratio, given by Eqn.~(\ref{eq:eaf}). 
If the attacker strategically deploys its node in the cloud service co-located with modified nodes, the cost is 0--20 USD/TB (AWS US East) including the case when the attacker sends transactions to multiple modified nodes across regions. 
On the one hand, the modified nodes are organically connected to the rest of the P2P network, with 20\% of internal AWS traffic (Appendix~\ref{appendix:software_isp}) and 80\% of external traffic to the internet. 
AWS dynamically charges external traffic starting from 90 USD/TB up to 10TB.
EAF is 13,827 for the first 10GB of non-AWS traffic.

\section{Attack simulation in the local P2P network}
\label{sec:attack_simulation}
To precision the attack mechanics, we simulate our proposed attack in the local network and compare it with the existing DoS attacks.

We set up a local P2P network, in which we deploy our modified node, attack the node, and then assess both txpool congestion and transaction inclusion in the
blocks. We base our attack on Yaish et al.'s publicly available
repository,\footnote{\url{https://github.com/AvivYaish/SpeculativeDoS}}
which allows for a direct comparison of our approach with the two
existing DoS attacks, namely 1) \emph{Baseline}, that is, sending valid
transactions with higher gas prices (naive eviction strategy), and 2)
\emph{MemPurge}~\cite{yaish2023speculative}, that is, sending future latent
(invalid) transactions by extending the DETER method~\cite{li2021deter}. 
We test all three attacks in our modified node. 

We set up one validator, 80 honest accounts, and a varying number $x$
of attack accounts, all of which interact with our 
node through the RPC endpoint. First, each honest account
sends 64 transactions to collectively fill up the txpool with honest transactions.
Each attacker then 
submit attack transactions to evict honest transactions. 
In our proposed attack, a attacker attempts to transfer an amount
higher than its current balance, and 
starting from nonce 0, which is lower than the current nonce -- i.e., 
these transactions are invalid. 
Each attack account then gradually increments the nonce 
to generate 32 invalid transactions (i.e., a total of $32x$ attack transactions).

\begin{figure*}[h]
  \begin{subfigure}{0.3\linewidth}
   \includegraphics[width=\textwidth]{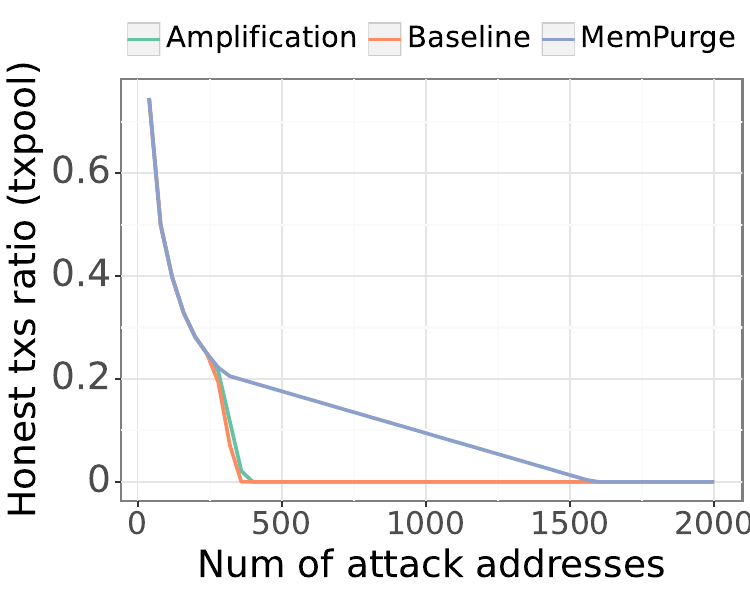}
  \end{subfigure}
  \begin{subfigure}{0.3\linewidth}
    \includegraphics[width=\textwidth]{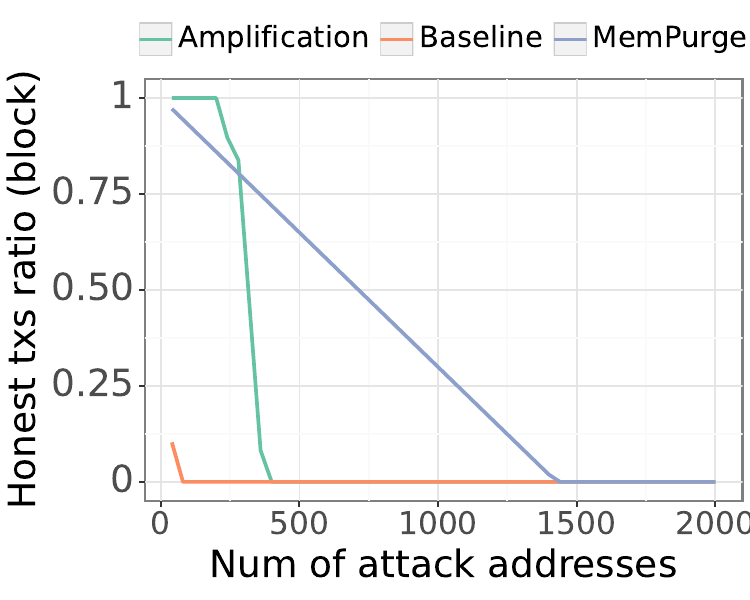}
  \end{subfigure}
  \begin{subfigure}{0.3\linewidth}
    \includegraphics[width=\textwidth]{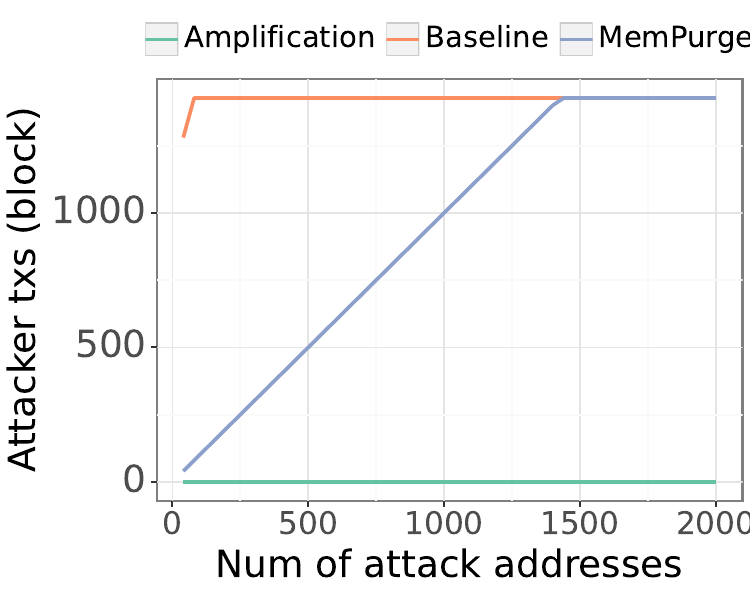}
  \end{subfigure}
  \caption{The ratio of honest transactions in the txpool (left), the ratio of honest transactions in the block (middle), the number of attack transactions in the block (right), over the number of attack accounts (x-axis).}
  \label{fig:attack_simulation} 
\end{figure*}

We measure 1) the ratio of honest transactions in the txpool, 2) the
ratio of honest transactions in the final block, and 3) the number of attack transactions in the block (i.e., the attack transaction cost). 
We calibrate the
number of malicious accounts controlled by the attacker to measure the
changes in those three metrics. Figure~\ref{fig:attack_simulation}
depicts the relationship between those three metrics ($y$-axis) and the
number of malicious accounts controlled by the attacker ($x$-axis).

Figure~\ref{fig:attack_simulation} (left) illustrates that our
proposed attack can evict as many honest transactions as
the Baseline attack. MemPurge requires even a larger number of
attack accounts to replace honest transactions, as the node pushes
back future transactions to the queue if the txpool is already
full.

Figures~\ref{fig:attack_simulation} (middle and right) show that our proposed attack
can exclude honest transactions not just in the txpool but block as well because the attacker's invalid transactions occupy
the txpool.
The delay (or the failure) in the transaction inclusion negatively impacts the security of the other layers (e.g., consensus)~\cite{gervais2015tampering, wahrstatter2023blockchain}. 
Furthermore, these invalid transactions are never
included in the block, resulting in zero attack transaction costs. MemPurge gradually
displaces honest transactions in the block, while its attacker pays
for one transaction per attack account. The Baseline attack replaces
honest transactions with a smaller number of attack accounts. Yet,
all the attack transactions are included in the block, resulting in a
significantly high(er) attack cost. 

In summary, our attack achieves the same level
of txpool/block eviction rate as the two existing DoS attacks but with
significantly lower (zero) attack transaction costs. 
This advantage allows the attacker to keep refreshing the txpool with its new invalid transactions, causing the modified node to continuously notify its neighbors, thereby incurring egress traffic loss---EDoS attack. 
In Appendix~\ref{appendix:attack_scinario}, we describe more detailed attack scenarios and roughly estimate the maximum traffic monthly costs incurred at the modified node to be 88,904 USD per node (8M USD in aggregate) when an attacker successfully saturates outgoing traffic with invalid transactions.

We open-source our attack code in our repository\footnote{\url{https://github.com/taro-tsuchiya/BlockchainAmplification}} and document the attack implementation details. 

\section{Estimating transaction validations}
\label{sec:estimate_val_time}
In this section, we conduct two experiments: 1) simulation, and 2) empirical measurement, focused on the latency benefits behind skipping certain transaction validation steps. 

The Geth node performs three sets of validations: 1) state-independent
validations, such as checking transaction size, parameter range, minimum
gas price, and signature; 2) state-dependent validations, such as
verifying account balance and nonce; and 3) txpool checks such as
the existence of the replicated transactions (``gas bump rule'' in
\S\ref{sec:bg}). The Geth node executes the above three validations
at the following three stages: \textit{Acccept}, \textit{Foward}, and
\textit{Update}.

\textit{Accept}: Each node validates transactions received from its peers or RPC endpoint and places them into a ``(non-executable) queue'': \href{https://github.com/ethereum/go-ethereum/blob/master/core/txpool/validation.go#L56}{\texttt{ValidateTransaction()}}, 
\href{https://github.com/ethereum/go-ethereum/blob/master/core/txpool/validation.go#L202}{\texttt{ValidateTransactionWithState()}}, 
\href{https://github.com/ethereum/go-ethereum/blob/master/core/txpool/legacypool/list.go#L303}{\texttt{Add()}}, 
\href{https://github.com/ethereum/go-ethereum/blob/master/core/txpool/legacypool/legacypool.go#L846)}{\texttt{enqueueTx()}}.

\textit{Forward}: The node re-validates/moves transactions into txpool, and starts forwarding them to peers: 
\href{https://github.com/ethereum/go-ethereum/blob/master/core/txpool/legacypool/legacypool.go#L1442}{\texttt{promoteExecutables()}},
\href{https://github.com/ethereum/go-ethereum/blob/master/core/txpool/legacypool/legacypool.go#L899}{\texttt{promoteTx()}}.

\textit{Update}: Upon the arrival of a new block, the node validates and removes transactions that are already included in the blockchain or become invalid due to a change in the blockchain state:
\href{https://github.com/ethereum/go-ethereum/blob/master/core/txpool/legacypool/legacypool.go#L1648}{\texttt{demoteUnexecutables()}}.

\subsection{Simulating validations}
\label{subsec:val_simulation}
We first create an isolated fork of the Ethereum blockchain within our local network (without impacting the live P2P network) to estimate the Geth validation processing time. 
We use a customized Geth client, integrated with the Pebble database.\footnote{Geth can run on either Pebble or LevelDB as its underlying database.} 
Our objective is to examine how different types of validations contribute
to the overall processing time and identify potential areas for
efficiency improvements. Specifically, we run two functions in
the first ``accept'' stage: {\texttt{ValidateTransaction()} and
{\texttt{ValidateTransactionWithState()} for 1,000 consecutive
blocks, starting from block number~18,140,000, in a total of~136,437
transactions. 
We conduct our experiment on a Ubuntu 22.04.2 LTS server, equipped with an AMD Ryzen Threadripper 3990X 64-Core Processor and a Corsair MP400 NVMe disk.

Our experimental results indicate that running the first validation stage takes roughly 1 millisecond per transaction.
In particular, state-independent validations constitute roughly $12\%$ of the total validation time, with an average duration of $0.08$ ms (std=$0.04$) per transaction while state-dependent validation requires $86\%$ (mean=$0.89$ ms, std=$1.19$) of the total time.
Within the state-dependent validation, nonce checks were the most time-consuming, accounting for about 82.2\% of the total validation time. 
This significant portion largely comes from 1) a large blockchain state size, 2) disk accesses, and 3) the inherent complexities of the Merkle Patricia Trie (MPT) structure used for managing the Ethereum blockchain state. 
Refer to the detailed analysis in Appendix~\ref{appendix:cpu_profile}.

\subsection{Empirically measuring validations}
\label{subsec:run_modified}
The previous experiment solely examines validation processing time,
omitting txpool or the timing of transaction arrival. We next
empirically measure the latency difference between modified and regular
nodes. Since our modified node may forward attackers’ invalid transactions,
we choose to run the node in the Ethereum testnet ``Holesky'' instead of the mainnet. Our modified node bypasses the first (``accept'')
and the second (``forward'') validation stages, but keeps the third
(``remove'') validation stage, because the node gets congested by past transactions and stops inserting new transactions. To validate
our implementation, we send one test invalid  transaction (insufficient balance and past nonce) through our node's
RPC and verify that this transaction appears in our modified nodes' txpool. 
To limit confounding factors, we alternate between running an
unmodified and a modified node on the \textit{same} machine. This allows us to
control for 1) the specs of the machine (Ubuntu 22.04.4 LTS,  
with an AMD EPYC 9124 16-Core processor) and 2) the network topology by connecting to a similar set of peers using the same public key.
To further eliminate the effect of peer connections, we increase
the peer limit from the default (50) to 300 to connect to as many active
peers as possible (based on the observation by Zhao et
al~\cite{zhao2024dethna} on the Goerli testnet). 
We sync the blockchain
ahead of time and run our node from April 4, 2024, 1 PM (UTC)
to July 12, 2024, 1 PM.
We switch the modified node and the regular node weekly or bi-weekly basis, resulting in 7 weeks for each node. 
We capture around 34 million transactions in total.

To compare performance, we monitor each txpool and record the
timestamp when the node finishes processing each transaction. We
use a metric called ``inclusion time''~\cite{wahrstatter2023blockchain}: the timestamp of blockchain
minus the first observation timestamp in our txpool. The longer the 
inclusion time, the faster the node discovers and processes transactions.
We exclude periods when the market behaves abnormally (e.g., due to attacks or testing); refer to Appendix~\ref{appendix:measuring_latency}. 

The median inclusion times are 10.44.
and 10.62 seconds for the regular node and for the modified node,
respectively. 
This result suggests the modified node delivers
transactions faster than the regular node as expected.
\footnote{We use the Mann–Whitney
U test---non-parametric test robust to outliers, to confirm
statistical significance.} 
Despite the long-term measurement, market conditions vary significantly on a daily basis (especially in the test net), making it difficult for us to make a fair comparison between nodes.
We discuss the limitations of this experiment in Appendix~\ref{appendix:measuring_latency}.

\section{Discussion}
\label{sec:discussion}
We first discuss if, despite the
potential attack damages, it could still be economically
rational for an operator to modify nodes. We then describe the limitations
and the mitigations against the attack.

\subsection{Cost-benefit analysis}
\label{subsec:cost_benefit_analysis}
We summarize the benefits and the costs for each player: an attacker, modified nodes, and regular nodes. 
The attacker can induce financial losses for modified nodes. 
For instance, an attacker economically competing with the entity operating modified nodes could use the attack to weaken the competition. 
The attacker can also use its invalid transactions to displace legitimate transactions, rendering the service unusable (i.e., causing a DoS) as shown in \S\ref{sec:attack_simulation}.

A modified node, by skipping validation, reduces the latency and 
provides a ``faster'' txpool view. 
To approximate the economic   
value of latency, we refer to recent work from Wahrst{\"a}tter et al~\cite{wahrstatter2023time} that examines the impact of block bid timing on MEV profits in Ethereum. 
We estimate that delaying one millisecond in block submission increases up to 0.00002 ETH (\$0.050) (see the detailed calculation in Appendix~\ref{appendix:time_to_money}.) 
We cross validate the number with two other studies: 0.0000065~ETH from Schwarz-Schilling et al.~\cite{schwarz2023time} and 0.000022~ETH from Babel et al~\cite{babel2024prof}, which align with our estimate.
If the MEV searchers or block proposers reduce latency by $x$ milliseconds, they could gain \$0.05$x$ extra per block, which is \textit{up to} \$10,800$x$ per month. 

While omitting validation checks leads to a possible reduction in
processing time by the order of milliseconds (\S\ref{subsec:val_simulation}), the marginal time savings may not necessarily justify
the potential damages from the blockchain amplification attack (i.e., up to \$89,000 per node in \S\ref{sec:attack_simulation}). 
Modified nodes also
expect to face degradation in the quality of service provided, due to
the increasing number of invalid transactions (as shown in \S\ref{sec:attack_simulation}). 
With this in mind, a
rationally economic node would still continue skipping validations as long
as the economic benefits (including revenue from users paying to connect to those services as discussed in \S\ref{sec:bg}) outweigh any financial losses incurred from the attack.

Regular nodes can expect to receive transactions faster by connecting to
the modified nodes but receive more invalid (less valid) transactions, which consume network capacity and CPU.
This fact indicates that our attack poses a threat to the security of \textit{all} the players in a P2P network. 
(Our attack does not cause congestion in these nodes' txpool, since regular nodes discard invalid transactions.)

\subsection{Limitations and mitigations}
\label{subsec:limitation}
A few key assumptions affect our estimation model. 
First, we assume that
the set of nodes we are connected to is a representative sample of
the overall population. However, if our node(s) is/are more likely to
connect to the node with many open connections, we might overestimate
the distribution of peer connections $g(x)$.
Second, our method of estimating the number of active peers requires the assumption of the network being static over time. 
We mitigate this concern by only including the nodes with a significant number of messages---stable nodes, reinforcing the validity of our assumption. 
Third, while our model assumes that the node transmits each invalid transaction once, in practice, the node can resend the same transaction multiple times, which could significantly increase the amplification factor.

\noindent\textbf{Possible mitigations.} Previous
work suggests a few countermeasures to prevent DoS/EDoS
attack~\cite{chowdhury2017economic}: 1) testing the client (Turing test, cryptographic puzzle) to exclude
non-human requests or 2) blacklisting users on abnormal
traffic. Those solutions do not directly translate to our context because 1)
a blockchain node is not interactive (i.e., a human would not be able
to answer challenges at the requested rate) and 2) node identity can be
easily spoofed. 
Unlike previous blockchain DoS attacks~\cite{li2021deter, yaish2023speculative,heo2023partitioning}, our identified vulnerability does not exist in the public client, thereby patching the current client does not resolve the issue. 
We devise three solutions tailored to our case.

First, one can enforce a stricter txpool policy (checking
balance, nonce, gas price bump) on modified nodes. Upon recovering more
stringent transaction checks in the modified nodes, attack vectors
diminish. However, as long as profits could potentially 
exceed losses on modified
nodes, we cannot expect the solution to be implemented.

Second, modified nodes could apply validations after they forward transactions to users. The idea is similar to the solutions
proposed by Das et al~\cite{das2021tuxedo} for delaying \textit{block}
validation in the context of the Verifier's Dilemma. After the modified node receives transactions, it 1)
directly relays them to users (privately) without validation, ensuring latency reduction, 2) validates transactions afterward, and 3) sends them to the
rest (or does not forward them at all). 
While this two-step approach does not negatively impacts the rest of the network, modified nodes still pay the cost of delivering invalid
transactions to users, creating a risk
for users to include invalid transactions.

Third, one could employ a reputation system on regular nodes (e.g.,the ISP blacklisting~\cite{clayton2009much} or the penalty mechanism in the Bitcoin P2P network~\cite{gervais2015tampering}).
If the (regular) nodes have a mechanism to disconnect from the modified nodes
that propagate many invalid transactions, the modified nodes would stop
hearing new transactions, getting isolated from the rest of the network.
However, the Ethereum community is generally hesitant to introduce the node reputation system for two reasons.
First, the system could produce false positives due to discrepancies in the synced status (i.e., invalid transactions being considered valid if the node lags behind) or differences in client implementations, leading to a partitioned network. 
Second, the system invites a cat-and-mouse game with adversaries that attempt to bypass/game the reputation system. 
See more discussions in the Geth community~\cite{geth_reputation,geth_reputation2}

\subsection{Ethics}
\label{sec:ethics}
Following Tang et al.~\cite{tangethical} on ethics of measuring blockchain P2P networks, our work consists of \textit{passive} measurements (i.e., not submitting transactions ourselves). 
We do collect the Ethereum node IP addresses, but this is already publicly available information. 
In an effort to perform responsible disclosure, we contacted the services that we discovered are running modified nodes and could be victims of this attack. 
At least one of those services appears to have fixed the vulnerability. 

\section{Related work}
\label{sec:related_work}
We discuss related work in P2P network measurements and network security.

\subsection{P2P network measurements}
\label{subsec:related_work_measurement}
Most measurements on blockchain P2P networks focus
on the network topology such as network
size~\cite{kim2018measuring, neudecker2016timing}, influential
nodes~\cite{li2021toposhot, maeng2021visualization}, connection stabilities/longevities~\cite{neudecker10characterization, daniel2019map},
and latency between peers~\cite{decker2013information,
gencer2018decentralization, tang2023strategic,
wahrstatter2023blockchain}. Some websites  
summarize those statistics in real-time for various networks: Bitcoin~\cite{bitnodes, coindance, bitcoinkit};
Ethereum~\cite{etherscan, ethernodes, reth_crawler}; 
Monero~\cite{monerohash}.

In particular, there has been extensive work on identifying
\textit{active} peer connections to further understand the propagation
flow over the network. The techniques developed include: utilizing
latency between peers~\cite{neudecker2016timing, daniel2019map},
isolation property and orphan transactions~\cite{delgado2019txprobe},
connection freshness~\cite{cao2020exploring}, and gas fee
change~\cite{li2021toposhot, zhao2024dethna}. However, these works are
either 1) chain-specific and do not apply to Ethereum, and/or 2) involve
active measurement (e.g., submitting transactions), which is expensive and could disrupt
the main network. 
Our proposed method in \S\ref{subsec:estimate_gx}
works on the Ethereum main network 1) without intervening in ongoing
activities and 2) without incurring any additional costs apart from
collecting data. 
Our method only considers the number of connections
and does not determine whether an active edge exists between two
specific nodes, making it harder to compare its accuracy with previous proposals.

Another body of measurement research 
focuses on characterizing nodes, e.g., centralized entities such as mining or relay nodes~\cite{miller2015discovering, daniel2019map, li2021toposhot, li2021deter}, AS/cloud services~\cite{feld2014analyzing, gencer2018decentralization, neudecker2016timing}, geo-distributions~\cite{neudecker10characterization, kim2018measuring, daniel2019map, maeng2021visualization, cao2020exploring, cortes2023unveiling}, and peer technical specifications (i.e., bandwidth, CPUs)~\cite{tang2023strategic, cortes2023unveiling}. 
Closest to our work is the literature
about the client (software) types -- for Bitcoin~\cite{neudecker10characterization}, Ethereum~\cite{kim2018measuring, cortes2023unveiling, grandjean2023ethereum}, and Zcash~\cite{daniel2019map}.
Those works mostly focus on the diversity of software and/or versions, with relatively scarce attention on the implementation and performance. 
To the best of our knowledge, there has not been any exploration of customized clients, a gap our study aims to bridge.

\subsection{P2P network security: DoS and spam}
\label{subsec:related_work_P2P_security}
Due to the rise of centralized entities---mining pools, exchanges, and relay nodes, Denial of Service (DoS) attacks on those services can have a disastrous impact on the overall network~\cite{vasek2014empirical,gervais2015tampering}. 
The most relevant work is by Li et al~\cite{li2021deter} (and its extension by Yaish et al~\cite{yaish2023speculative}), which 
sends transactions with a future nonce to evict legitimate transactions in the txpool.
Further, Wang et al~\cite{wang2023understanding} develop a fuzzer to automatically discover txpool vulnerabilities.  
Heo et al~\cite{heo2023partitioning} illustrate that attackers can delay honest transaction execution by sending invalid transactions. 
Zhou et al~\cite{zhou2021a2mm} document the spamming behavior of MEV bots as a computational overhead to the network. 
Our paper describes a new blockchain DoS attack, exploiting the lack of transaction validations and empirically derives the amplification factors. 
We also reference some metrics from non-blockchain DoS literature such as amplified DoS and EDoS attacks from Kumar~\cite{kumar2007smurf} and Wang et al.~\cite{wang2016abusing}, respectively.

\section{Conclusion}
\label{sec:conclusion}
Latency reduction in blockchain P2P networks such as Ethereum can create economic value as a form of extractable value (MEV/BEV); as a result, 
some entities modify software clients (``modified nodes'') to
shorten the transaction validation process. Unfortunately, these modifications enable a Blockchain Amplification Attack. 
We formalize this attack, illustrate its practicality from similar attack instances observed in the wild, and empirically measure and simulate its impact. Specifically, we show that every byte a single attacker sends can result in 3,600 bytes traveling on the network; likewise, every dollar the attacker spends on the attack can result in \$13,800 worth of economic loss. 
We also quantify the benefits of skipping validation based on our simulations and measurements, highlighting the economic trade-off centralized entities face, when deciding whether to accept the risk of attack. 
We conclude by noting that a lack of transaction validation not only increases traffic costs on the modified nodes, but also introduces invalid transactions and degrades the user experience. 
Our models provide a foundation for explaining how invalid transactions propagates through the network and pose a threat to the overall blockchain ecosystem. 

\begin{acks}
This research was partially supported by Carnegie Mellon CyLab’s
Secure Blockchain Initiative and by the Nakajima Foundation.
\end{acks}

\bibliographystyle{ACM-Reference-Format}
\bibliography{ref}


\begin{thebibliography}{61}


\ifx \showCODEN    \undefined \def \showCODEN     #1{\unskip}     \fi
\ifx \showDOI      \undefined \def \showDOI       #1{#1}\fi
\ifx \showISBNx    \undefined \def \showISBNx     #1{\unskip}     \fi
\ifx \showISBNxiii \undefined \def \showISBNxiii  #1{\unskip}     \fi
\ifx \showISSN     \undefined \def \showISSN      #1{\unskip}     \fi
\ifx \showLCCN     \undefined \def \showLCCN      #1{\unskip}     \fi
\ifx \shownote     \undefined \def \shownote      #1{#1}          \fi
\ifx \showarticletitle \undefined \def \showarticletitle #1{#1}   \fi
\ifx \showURL      \undefined \def \showURL       {\relax}        \fi
\providecommand\bibfield[2]{#2}
\providecommand\bibinfo[2]{#2}
\providecommand\natexlab[1]{#1}
\providecommand\showeprint[2][]{arXiv:#2}

\bibitem[mem(2023)]%
        {mempoolguru}
 \bibinfo{year}{2023}\natexlab{}.
\newblock \bibinfo{title}{Mempool Guru}.
\newblock \bibinfo{howpublished}{\url{https://mempool.guru/}}.
\newblock
\newblock
\shownote{Last accessed Nov. 29th, 2023}.


\bibitem[Babel et~al\mbox{.}(2024)]%
        {babel2024prof}
\bibfield{author}{\bibinfo{person}{Kushal Babel}, \bibinfo{person}{Nerla
  Jean-Louis}, \bibinfo{person}{Yan Ji}, \bibinfo{person}{Ujval Misra},
  \bibinfo{person}{Mahimna Kelkar}, \bibinfo{person}{Kosala~Yapa
  Mudiyanselage}, \bibinfo{person}{Andrew Miller}, {and} \bibinfo{person}{Ari
  Juels}.} \bibinfo{year}{2024}\natexlab{}.
\newblock \showarticletitle{PROF: Protected Order Flow in a Profit-Seeking
  World}.
\newblock \bibinfo{journal}{\emph{arXiv preprint arXiv:2408.02303}}
  (\bibinfo{year}{2024}).
\newblock


\bibitem[bitnodes.io(2023)]%
        {bitnodes}
\bibfield{author}{\bibinfo{person}{bitnodes.io}.}
  \bibinfo{year}{2023}\natexlab{}.
\newblock \bibinfo{title}{Reachable bitcoin nodes}.
\newblock \bibinfo{howpublished}{\url{https://bitnodes.io/}}.
\newblock
\newblock
\shownote{Accessed Jul. 20th, 2023}.


\bibitem[Bloxroute(2019)]%
        {blx_bdn}
\bibfield{author}{\bibinfo{person}{Bloxroute}.}
  \bibinfo{year}{2019}\natexlab{}.
\newblock \bibinfo{title}{How bloXroute Achieves Its Performance}.
\newblock
  \bibinfo{howpublished}{\url{https://medium.com/bloxroute/how-bloxroute-achieves-its-performance-c408de842e67}}.
\newblock
\newblock
\shownote{Accessed Dec. 18th, 2023}.


\bibitem[Cao et~al\mbox{.}(2020)]%
        {cao2020exploring}
\bibfield{author}{\bibinfo{person}{Tong Cao}, \bibinfo{person}{Jiangshan Yu},
  \bibinfo{person}{J{\'e}r{\'e}mie Decouchant}, \bibinfo{person}{Xiapu Luo},
  {and} \bibinfo{person}{Paulo Verissimo}.} \bibinfo{year}{2020}\natexlab{}.
\newblock \showarticletitle{Exploring the monero peer-to-peer network}. In
  \bibinfo{booktitle}{\emph{Financial Cryptography and Data Security: 24th
  International Conference, FC 2020, Kota Kinabalu, Malaysia, February 10--14,
  2020 Revised Selected Papers 24}}. Springer, \bibinfo{pages}{578--594}.
\newblock


\bibitem[Chowdhury et~al\mbox{.}(2017)]%
        {chowdhury2017economic}
\bibfield{author}{\bibinfo{person}{Fahad~Zaman Chowdhury},
  \bibinfo{person}{Laiha Binti~Mat Kiah}, \bibinfo{person}{MA~Manazir Ahsan},
  {and} \bibinfo{person}{Mohd Yamani Idna~Bin Idris}.}
  \bibinfo{year}{2017}\natexlab{}.
\newblock \showarticletitle{Economic denial of sustainability (EDoS) mitigation
  approaches in cloud: Analysis and open challenges}. In
  \bibinfo{booktitle}{\emph{2017 International Conference on Electrical
  Engineering and Computer Science (ICECOS)}}. IEEE, \bibinfo{pages}{206--211}.
\newblock


\bibitem[Clayton(2009)]%
        {clayton2009much}
\bibfield{author}{\bibinfo{person}{Richard Clayton}.}
  \bibinfo{year}{2009}\natexlab{}.
\newblock \showarticletitle{How much did shutting down McColo help}.
\newblock \bibinfo{journal}{\emph{Proc. of 6th CEAS}} (\bibinfo{year}{2009}).
\newblock


\bibitem[coin.dance(2023)]%
        {coindance}
\bibfield{author}{\bibinfo{person}{coin.dance}.}
  \bibinfo{year}{2023}\natexlab{}.
\newblock \bibinfo{title}{Bitcoin Nodes Summary}.
\newblock \bibinfo{howpublished}{\url{https://coin.dance/nodes}}.
\newblock
\newblock
\shownote{Accessed Jul. 26th, 2023}.


\bibitem[Cortes-Goicoechea et~al\mbox{.}(2023)]%
        {cortes2023unveiling}
\bibfield{author}{\bibinfo{person}{Mikel Cortes-Goicoechea},
  \bibinfo{person}{Tarun Mohandas-Daryanani}, \bibinfo{person}{Jose~Luis
  Munoz-Tapia}, {and} \bibinfo{person}{Leonardo Bautista-Gomez}.}
  \bibinfo{year}{2023}\natexlab{}.
\newblock \showarticletitle{Unveiling Ethereum's Hidden Centralization
  Incentives: Does Connectivity Impact Performance?}
\newblock \bibinfo{journal}{\emph{arXiv preprint arXiv:2309.13329}}
  (\bibinfo{year}{2023}).
\newblock


\bibitem[Daian et~al\mbox{.}(2020)]%
        {daian2020flash}
\bibfield{author}{\bibinfo{person}{Philip Daian}, \bibinfo{person}{Steven
  Goldfeder}, \bibinfo{person}{Tyler Kell}, \bibinfo{person}{Yunqi Li},
  \bibinfo{person}{Xueyuan Zhao}, \bibinfo{person}{Iddo Bentov},
  \bibinfo{person}{Lorenz Breidenbach}, {and} \bibinfo{person}{Ari Juels}.}
  \bibinfo{year}{2020}\natexlab{}.
\newblock \showarticletitle{Flash boys 2.0: Frontrunning in decentralized
  exchanges, miner extractable value, and consensus instability}. In
  \bibinfo{booktitle}{\emph{2020 IEEE Symposium on Security and Privacy (SP)}}.
  IEEE, \bibinfo{pages}{910--927}.
\newblock


\bibitem[Daniel et~al\mbox{.}(2019)]%
        {daniel2019map}
\bibfield{author}{\bibinfo{person}{Erik Daniel}, \bibinfo{person}{Elias
  Rohrer}, {and} \bibinfo{person}{Florian Tschorsch}.}
  \bibinfo{year}{2019}\natexlab{}.
\newblock \showarticletitle{Map-z: Exposing the zcash network in times of
  transition}. In \bibinfo{booktitle}{\emph{2019 IEEE 44th Conference on Local
  Computer Networks (LCN)}}. IEEE, \bibinfo{pages}{84--92}.
\newblock


\bibitem[Das et~al\mbox{.}(2021)]%
        {das2021tuxedo}
\bibfield{author}{\bibinfo{person}{Sourav Das}, \bibinfo{person}{Nitin
  Awathare}, \bibinfo{person}{Ling Ren}, \bibinfo{person}{Vinay~J Ribeiro},
  {and} \bibinfo{person}{Umesh Bellur}.} \bibinfo{year}{2021}\natexlab{}.
\newblock \showarticletitle{Tuxedo: maximizing smart contract computation in
  PoW blockchains}.
\newblock \bibinfo{journal}{\emph{Proceedings of the ACM on Measurement and
  Analysis of Computing Systems}} \bibinfo{volume}{5}, \bibinfo{number}{3}
  (\bibinfo{year}{2021}), \bibinfo{pages}{1--30}.
\newblock


\bibitem[Decker and Wattenhofer(2013)]%
        {decker2013information}
\bibfield{author}{\bibinfo{person}{Christian Decker} {and}
  \bibinfo{person}{Roger Wattenhofer}.} \bibinfo{year}{2013}\natexlab{}.
\newblock \showarticletitle{Information propagation in the bitcoin network}. In
  \bibinfo{booktitle}{\emph{IEEE P2P 2013 Proceedings}}. IEEE,
  \bibinfo{pages}{1--10}.
\newblock


\bibitem[Delgado-Segura et~al\mbox{.}(2019)]%
        {delgado2019txprobe}
\bibfield{author}{\bibinfo{person}{Sergi Delgado-Segura},
  \bibinfo{person}{Surya Bakshi}, \bibinfo{person}{Cristina
  P{\'e}rez-Sol{\`a}}, \bibinfo{person}{James Litton}, \bibinfo{person}{Andrew
  Pachulski}, \bibinfo{person}{Andrew Miller}, {and} \bibinfo{person}{Bobby
  Bhattacharjee}.} \bibinfo{year}{2019}\natexlab{}.
\newblock \showarticletitle{Txprobe: Discovering bitcoin’s network topology
  using orphan transactions}. In \bibinfo{booktitle}{\emph{Financial
  Cryptography and Data Security: 23rd International Conference, FC 2019,
  Frigate Bay, St. Kitts and Nevis, February 18--22, 2019, Revised Selected
  Papers 23}}. Springer, \bibinfo{pages}{550--566}.
\newblock


\bibitem[erigon(2024)]%
        {erigon_propagation_policy_analysis}
\bibfield{author}{\bibinfo{person}{erigon}.} \bibinfo{year}{2024}\natexlab{}.
\newblock \bibinfo{title}{txpool: limit transactions outgoing messages (\#8271)
  Conversation}.
\newblock
  \bibinfo{howpublished}{\url{https://github.com/ledgerwatch/erigon/pull/8742}}.
\newblock
\newblock
\shownote{Accessed Jan. 30th, 2024}.


\bibitem[Ethereum(2023a)]%
        {ethereum_wire}
\bibfield{author}{\bibinfo{person}{Ethereum}.}
  \bibinfo{year}{2023}\natexlab{a}.
\newblock \bibinfo{title}{Ethereum devp2p Documentation (Ethereum Wire
  Protocol}.
\newblock
  \bibinfo{howpublished}{\url{https://github.com/ethereum/devp2p/blob/master/caps/eth.md}}.
\newblock
\newblock
\shownote{Accessed Dec. 3rd, 2023}.


\bibitem[Ethereum(2023b)]%
        {ethereum_discv4}
\bibfield{author}{\bibinfo{person}{Ethereum}.}
  \bibinfo{year}{2023}\natexlab{b}.
\newblock \bibinfo{title}{Ethereum devp2p Documentation (Node Discovery
  Protocol)}.
\newblock
  \bibinfo{howpublished}{\url{https://github.com/ethereum/devp2p/blob/master/discv4.md}}.
\newblock
\newblock
\shownote{Accessed Dec. 3rd, 2023}.


\bibitem[Ethernodes.org(2023)]%
        {ethernodes}
\bibfield{author}{\bibinfo{person}{Ethernodes.org}.}
  \bibinfo{year}{2023}\natexlab{}.
\newblock \bibinfo{title}{Ethereum Mainnet Statistics}.
\newblock \bibinfo{howpublished}{\url{https://ethernodes.org/}}.
\newblock
\newblock
\shownote{Accessed Jul. 20th, 2023}.


\bibitem[Etherscan.io(2023)]%
        {etherscan}
\bibfield{author}{\bibinfo{person}{Etherscan.io}.}
  \bibinfo{year}{2023}\natexlab{}.
\newblock \bibinfo{title}{Ethereum Node Tracker}.
\newblock \bibinfo{howpublished}{\url{https://etherscan.io/nodetracker}}.
\newblock
\newblock
\shownote{Accessed Nov. 28th, 2023}.


\bibitem[Eyal and Sirer(2018)]%
        {eyal2018majority}
\bibfield{author}{\bibinfo{person}{Ittay Eyal} {and}
  \bibinfo{person}{Emin~G{\"u}n Sirer}.} \bibinfo{year}{2018}\natexlab{}.
\newblock \showarticletitle{Majority is not enough: Bitcoin mining is
  vulnerable}.
\newblock \bibinfo{journal}{\emph{Commun. ACM}} \bibinfo{volume}{61},
  \bibinfo{number}{7} (\bibinfo{year}{2018}), \bibinfo{pages}{95--102}.
\newblock


\bibitem[Feld et~al\mbox{.}(2014)]%
        {feld2014analyzing}
\bibfield{author}{\bibinfo{person}{Sebastian Feld}, \bibinfo{person}{Mirco
  Sch{\"o}nfeld}, {and} \bibinfo{person}{Martin Werner}.}
  \bibinfo{year}{2014}\natexlab{}.
\newblock \showarticletitle{Analyzing the Deployment of Bitcoin's P2P Network
  under an AS-level Perspective}.
\newblock \bibinfo{journal}{\emph{Procedia Computer Science}}
  \bibinfo{volume}{32} (\bibinfo{year}{2014}), \bibinfo{pages}{1121--1126}.
\newblock


\bibitem[Gencer et~al\mbox{.}(2018)]%
        {gencer2018decentralization}
\bibfield{author}{\bibinfo{person}{Adem~Efe Gencer}, \bibinfo{person}{Soumya
  Basu}, \bibinfo{person}{Ittay Eyal}, \bibinfo{person}{Robbert Van~Renesse},
  {and} \bibinfo{person}{Emin~G{\"u}n Sirer}.} \bibinfo{year}{2018}\natexlab{}.
\newblock \showarticletitle{Decentralization in bitcoin and ethereum networks}.
  In \bibinfo{booktitle}{\emph{Financial Cryptography and Data Security: 22nd
  International Conference, FC 2018, Nieuwpoort, Cura{\c{c}}ao, February
  26--March 2, 2018, Revised Selected Papers 22}}. Springer,
  \bibinfo{pages}{439--457}.
\newblock


\bibitem[Gervais et~al\mbox{.}(2016)]%
        {gervais2016security}
\bibfield{author}{\bibinfo{person}{Arthur Gervais}, \bibinfo{person}{Ghassan~O
  Karame}, \bibinfo{person}{Karl W{\"u}st}, \bibinfo{person}{Vasileios
  Glykantzis}, \bibinfo{person}{Hubert Ritzdorf}, {and} \bibinfo{person}{Srdjan
  Capkun}.} \bibinfo{year}{2016}\natexlab{}.
\newblock \showarticletitle{On the security and performance of proof of work
  blockchains}. In \bibinfo{booktitle}{\emph{Proceedings of the 2016 ACM SIGSAC
  conference on computer and communications security}}. \bibinfo{pages}{3--16}.
\newblock


\bibitem[Gervais et~al\mbox{.}(2015)]%
        {gervais2015tampering}
\bibfield{author}{\bibinfo{person}{Arthur Gervais}, \bibinfo{person}{Hubert
  Ritzdorf}, \bibinfo{person}{Ghassan~O Karame}, {and} \bibinfo{person}{Srdjan
  Capkun}.} \bibinfo{year}{2015}\natexlab{}.
\newblock \showarticletitle{Tampering with the delivery of blocks and
  transactions in bitcoin}. In \bibinfo{booktitle}{\emph{Proceedings of the
  22nd ACM SIGSAC Conference on Computer and Communications Security}}.
  \bibinfo{pages}{692--705}.
\newblock


\bibitem[Geth(2024a)]%
        {geth_reputation}
\bibfield{author}{\bibinfo{person}{Geth}.} \bibinfo{year}{2024}\natexlab{a}.
\newblock \bibinfo{title}{Connection Slot Exhaustion with Passive Nodes
  \#29329}.
\newblock
  \bibinfo{howpublished}{\url{https://github.com/ethereum/go-ethereum/issues/29329}}.
\newblock
\newblock
\shownote{Accessed Apr. 18th, 2024}.


\bibitem[Geth(2024b)]%
        {geth_reputation2}
\bibfield{author}{\bibinfo{person}{Geth}.} \bibinfo{year}{2024}\natexlab{b}.
\newblock \bibinfo{title}{Drop Selfish Peers in Transaction Sharing \#29327}.
\newblock
  \bibinfo{howpublished}{\url{https://github.com/ethereum/go-ethereum/issues/29327/\#issuecomment-2017100792}}.
\newblock
\newblock
\shownote{Accessed Dec. 11th, 2024}.


\bibitem[Grandjean et~al\mbox{.}(2023)]%
        {grandjean2023ethereum}
\bibfield{author}{\bibinfo{person}{Dominic Grandjean}, \bibinfo{person}{Lioba
  Heimbach}, {and} \bibinfo{person}{Roger Wattenhofer}.}
  \bibinfo{year}{2023}\natexlab{}.
\newblock \showarticletitle{Ethereum Proof-of-Stake Consensus Layer:
  Participation and Decentralization}.
\newblock \bibinfo{journal}{\emph{arXiv preprint arXiv:2306.10777}}
  (\bibinfo{year}{2023}).
\newblock


\bibitem[Hager(2023)]%
        {flashbot_md_data}
\bibfield{author}{\bibinfo{person}{Chris Hager}.}
  \bibinfo{year}{2023}\natexlab{}.
\newblock \bibinfo{title}{Mempool Dumpster}.
\newblock
  \bibinfo{howpublished}{\url{https://mempool-dumpster.flashbots.net/index.html}}.
\newblock
\newblock
\shownote{Last accessed Nov. 29th, 2023}.


\bibitem[He et~al\mbox{.}(2024)]%
        {he2024nurgle}
\bibfield{author}{\bibinfo{person}{Zheyuan He}, \bibinfo{person}{Zihao Li},
  \bibinfo{person}{Ao Qiao}, \bibinfo{person}{Xiapu Luo},
  \bibinfo{person}{Xiaosong Zhang}, \bibinfo{person}{Ting Chen},
  \bibinfo{person}{Shuwei Song}, \bibinfo{person}{Dijun Liu}, {and}
  \bibinfo{person}{Weina Niu}.} \bibinfo{year}{2024}\natexlab{}.
\newblock \showarticletitle{NURGLE: Exacerbating Resource Consumption in
  Blockchain State Storage via MPT Manipulation}. In
  \bibinfo{booktitle}{\emph{2024 IEEE Symposium on Security and Privacy (SP)}}.
  IEEE.
\newblock


\bibitem[Heilman et~al\mbox{.}(2015)]%
        {heilman2015eclipse}
\bibfield{author}{\bibinfo{person}{Ethan Heilman}, \bibinfo{person}{Alison
  Kendler}, \bibinfo{person}{Aviv Zohar}, {and} \bibinfo{person}{Sharon
  Goldberg}.} \bibinfo{year}{2015}\natexlab{}.
\newblock \showarticletitle{Eclipse attacks on
  $\{$Bitcoin’s$\}$$\{$peer-to-peer$\}$ network}. In
  \bibinfo{booktitle}{\emph{24th USENIX security symposium (USENIX security
  15)}}. \bibinfo{pages}{129--144}.
\newblock


\bibitem[Heo et~al\mbox{.}(2023)]%
        {heo2023partitioning}
\bibfield{author}{\bibinfo{person}{Hwanjo Heo}, \bibinfo{person}{Seungwon Woo},
  \bibinfo{person}{Taeung Yoon}, \bibinfo{person}{Min~Suk Kang}, {and}
  \bibinfo{person}{Seungwon Shin}.} \bibinfo{year}{2023}\natexlab{}.
\newblock \showarticletitle{Partitioning Ethereum without Eclipsing It.}. In
  \bibinfo{booktitle}{\emph{NDSS}}.
\newblock


\bibitem[Hoff(2008)]%
        {hoff2008edos}
\bibfield{author}{\bibinfo{person}{Christopher Hoff}.}
  \bibinfo{year}{2008}\natexlab{}.
\newblock \bibinfo{title}{Cloud Computing Security: From DDoS (Distributed
  Denial Of Service) to EDoS (Economic Denial of Sustainability)}.
\newblock
  \bibinfo{howpublished}{\url{https://rationalsecurity.typepad.com/blog/2008/11/cloud-computing-security-from-ddos-distributed-denial-of-service-to-edos-economic-denial-of-sustaina.html}}.
\newblock


\bibitem[Karame et~al\mbox{.}(2012)]%
        {karame2012double}
\bibfield{author}{\bibinfo{person}{Ghassan~O Karame}, \bibinfo{person}{Elli
  Androulaki}, {and} \bibinfo{person}{Srdjan Capkun}.}
  \bibinfo{year}{2012}\natexlab{}.
\newblock \showarticletitle{Double-spending fast payments in bitcoin}. In
  \bibinfo{booktitle}{\emph{Proceedings of the 2012 ACM conference on Computer
  and communications security}}. \bibinfo{pages}{906--917}.
\newblock


\bibitem[Kim et~al\mbox{.}(2018)]%
        {kim2018measuring}
\bibfield{author}{\bibinfo{person}{Seoung~Kyun Kim}, \bibinfo{person}{Zane Ma},
  \bibinfo{person}{Siddharth Murali}, \bibinfo{person}{Joshua Mason},
  \bibinfo{person}{Andrew Miller}, {and} \bibinfo{person}{Michael Bailey}.}
  \bibinfo{year}{2018}\natexlab{}.
\newblock \showarticletitle{Measuring ethereum network peers}. In
  \bibinfo{booktitle}{\emph{Proceedings of the Internet Measurement Conference
  2018}}. \bibinfo{pages}{91--104}.
\newblock


\bibitem[Klarman et~al\mbox{.}(2018)]%
        {klarman2018bloxroute}
\bibfield{author}{\bibinfo{person}{Uri Klarman}, \bibinfo{person}{Soumya Basu},
  \bibinfo{person}{Aleksandar Kuzmanovic}, {and} \bibinfo{person}{Emin~G{\"u}n
  Sirer}.} \bibinfo{year}{2018}\natexlab{}.
\newblock \showarticletitle{bloxroute: A scalable trustless blockchain
  distribution network whitepaper}.
\newblock \bibinfo{journal}{\emph{IEEE Internet of Things Journal}}
  (\bibinfo{year}{2018}).
\newblock


\bibitem[Kumar(2007)]%
        {kumar2007smurf}
\bibfield{author}{\bibinfo{person}{Sanjeev Kumar}.}
  \bibinfo{year}{2007}\natexlab{}.
\newblock \showarticletitle{Smurf-based distributed denial of service (ddos)
  attack amplification in internet}. In \bibinfo{booktitle}{\emph{Second
  International Conference on Internet Monitoring and Protection (ICIMP
  2007)}}. IEEE, \bibinfo{pages}{25--25}.
\newblock


\bibitem[Lewis(2014)]%
        {lewis2014flash}
\bibfield{author}{\bibinfo{person}{Michael Lewis}.}
  \bibinfo{year}{2014}\natexlab{}.
\newblock \bibinfo{booktitle}{\emph{Flash boys: a Wall Street revolt}}.
\newblock \bibinfo{publisher}{WW Norton \& Company}.
\newblock


\bibitem[Li et~al\mbox{.}(2021a)]%
        {li2021toposhot}
\bibfield{author}{\bibinfo{person}{Kai Li}, \bibinfo{person}{Yuzhe Tang},
  \bibinfo{person}{Jiaqi Chen}, \bibinfo{person}{Yibo Wang}, {and}
  \bibinfo{person}{Xianghong Liu}.} \bibinfo{year}{2021}\natexlab{a}.
\newblock \showarticletitle{TopoShot: uncovering Ethereum's network topology
  leveraging replacement transactions}. In
  \bibinfo{booktitle}{\emph{Proceedings of the 21st ACM Internet Measurement
  Conference}}. \bibinfo{pages}{302--319}.
\newblock


\bibitem[Li et~al\mbox{.}(2021b)]%
        {li2021deter}
\bibfield{author}{\bibinfo{person}{Kai Li}, \bibinfo{person}{Yibo Wang}, {and}
  \bibinfo{person}{Yuzhe Tang}.} \bibinfo{year}{2021}\natexlab{b}.
\newblock \showarticletitle{Deter: Denial of ethereum txpool services}. In
  \bibinfo{booktitle}{\emph{Proceedings of the 2021 ACM SIGSAC Conference on
  Computer and Communications Security}}. \bibinfo{pages}{1645--1667}.
\newblock


\bibitem[Luu et~al\mbox{.}(2015)]%
        {luu2015demystifying}
\bibfield{author}{\bibinfo{person}{Loi Luu}, \bibinfo{person}{Jason Teutsch},
  \bibinfo{person}{Raghav Kulkarni}, {and} \bibinfo{person}{Prateek Saxena}.}
  \bibinfo{year}{2015}\natexlab{}.
\newblock \showarticletitle{Demystifying incentives in the consensus computer}.
  In \bibinfo{booktitle}{\emph{Proceedings of the 22Nd acm sigsac conference on
  computer and communications security}}. \bibinfo{pages}{706--719}.
\newblock


\bibitem[Maeng et~al\mbox{.}(2021)]%
        {maeng2021visualization}
\bibfield{author}{\bibinfo{person}{Soohoon Maeng}, \bibinfo{person}{Meryam
  Essaid}, \bibinfo{person}{Changhyun Lee}, \bibinfo{person}{Sejin Park}, {and}
  \bibinfo{person}{Hongteak Ju}.} \bibinfo{year}{2021}\natexlab{}.
\newblock \showarticletitle{Visualization of Ethereum P2P network topology and
  peer properties}.
\newblock \bibinfo{journal}{\emph{International Journal of Network Management}}
  \bibinfo{volume}{31}, \bibinfo{number}{6} (\bibinfo{year}{2021}),
  \bibinfo{pages}{e2175}.
\newblock


\bibitem[Maymounkov and Mazieres(2002)]%
        {maymounkov2002kademlia}
\bibfield{author}{\bibinfo{person}{Petar Maymounkov} {and}
  \bibinfo{person}{David Mazieres}.} \bibinfo{year}{2002}\natexlab{}.
\newblock \showarticletitle{Kademlia: A peer-to-peer information system based
  on the xor metric}. In \bibinfo{booktitle}{\emph{International Workshop on
  Peer-to-Peer Systems}}. Springer, \bibinfo{pages}{53--65}.
\newblock


\bibitem[Mazza(2023)]%
        {reth_crawler}
\bibfield{author}{\bibinfo{person}{Alessandro Mazza}.}
  \bibinfo{year}{2023}\natexlab{}.
\newblock \bibinfo{title}{Reth-crawler}.
\newblock \bibinfo{howpublished}{\url{https://etherclients.com/}}.
\newblock
\newblock
\shownote{Accessed Dec. 7th, 2023}.


\bibitem[Miller et~al\mbox{.}(2015)]%
        {miller2015discovering}
\bibfield{author}{\bibinfo{person}{Andrew Miller}, \bibinfo{person}{James
  Litton}, \bibinfo{person}{Andrew Pachulski}, \bibinfo{person}{Neal Gupta},
  \bibinfo{person}{Dave Levin}, \bibinfo{person}{Neil Spring},
  \bibinfo{person}{Bobby Bhattacharjee}, {et~al\mbox{.}}}
  \bibinfo{year}{2015}\natexlab{}.
\newblock \showarticletitle{Discovering bitcoin’s public topology and
  influential nodes}.
\newblock \bibinfo{journal}{\emph{et al}} (\bibinfo{year}{2015}).
\newblock


\bibitem[MoneroHash.com(2023)]%
        {monerohash}
\bibfield{author}{\bibinfo{person}{MoneroHash.com}.}
  \bibinfo{year}{2023}\natexlab{}.
\newblock \bibinfo{howpublished}{\url{https://monerohash.com/}}.
\newblock
\newblock
\shownote{Accessed Jul. 25th, 2023}.


\bibitem[Neudecker(2019)]%
        {neudecker10characterization}
\bibfield{author}{\bibinfo{person}{T Neudecker}.}
  \bibinfo{year}{2019}\natexlab{}.
\newblock \showarticletitle{Characterization of the bitcoin peer-to-peer
  network (2015--2018)(2019)}.
\newblock \bibinfo{journal}{\emph{DOI: https://doi. org/10.5445/IR/1000091933}}
  (\bibinfo{year}{2019}).
\newblock


\bibitem[Neudecker et~al\mbox{.}(2016)]%
        {neudecker2016timing}
\bibfield{author}{\bibinfo{person}{Till Neudecker}, \bibinfo{person}{Philipp
  Andelfinger}, {and} \bibinfo{person}{Hannes Hartenstein}.}
  \bibinfo{year}{2016}\natexlab{}.
\newblock \showarticletitle{Timing analysis for inferring the topology of the
  bitcoin peer-to-peer network}. In \bibinfo{booktitle}{\emph{2016 Intl IEEE
  Conferences on Ubiquitous Intelligence \& Computing, Advanced and Trusted
  Computing, Scalable Computing and Communications, Cloud and Big Data
  Computing, Internet of People, and Smart World Congress
  (UIC/ATC/ScalCom/CBDCom/IoP/SmartWorld)}}. IEEE, \bibinfo{pages}{358--367}.
\newblock


\bibitem[Qin et~al\mbox{.}(2022)]%
        {qin2022quantifying}
\bibfield{author}{\bibinfo{person}{Kaihua Qin}, \bibinfo{person}{Liyi Zhou},
  {and} \bibinfo{person}{Arthur Gervais}.} \bibinfo{year}{2022}\natexlab{}.
\newblock \showarticletitle{Quantifying blockchain extractable value: How dark
  is the forest?}. In \bibinfo{booktitle}{\emph{2022 IEEE Symposium on Security
  and Privacy (SP)}}. IEEE, \bibinfo{pages}{198--214}.
\newblock


\bibitem[Schwarz-Schilling et~al\mbox{.}(2023)]%
        {schwarz2023time}
\bibfield{author}{\bibinfo{person}{Caspar Schwarz-Schilling},
  \bibinfo{person}{Fahad Saleh}, \bibinfo{person}{Thomas Thiery},
  \bibinfo{person}{Jennifer Pan}, \bibinfo{person}{Nihar Shah}, {and}
  \bibinfo{person}{Barnab{\'e} Monnot}.} \bibinfo{year}{2023}\natexlab{}.
\newblock \showarticletitle{Time is Money: Strategic Timing Games in
  Proof-of-Stake Protocols}.
\newblock \bibinfo{journal}{\emph{arXiv preprint arXiv:2305.09032}}
  (\bibinfo{year}{2023}).
\newblock


\bibitem[Systems and KASTEL(2023)]%
        {bitcoinkit}
\bibfield{author}{\bibinfo{person}{Decentralized Systems} {and}
  \bibinfo{person}{Network Services Research~Group KASTEL}.}
  \bibinfo{year}{2023}\natexlab{}.
\newblock \bibinfo{title}{Bitcoin Monitoring}.
\newblock
  \bibinfo{howpublished}{\url{https://www.dsn.kastel.kit.edu/bitcoin/}}.
\newblock
\newblock
\shownote{Accessed Jul. 27th, 2023}.


\bibitem[Tang et~al\mbox{.}(2023a)]%
        {tang2023strategic}
\bibfield{author}{\bibinfo{person}{Weizhao Tang}, \bibinfo{person}{Lucianna
  Kiffer}, \bibinfo{person}{Giulia Fanti}, {and} \bibinfo{person}{Ari Juels}.}
  \bibinfo{year}{2023}\natexlab{a}.
\newblock \showarticletitle{Strategic Latency Reduction in Blockchain
  Peer-to-Peer Networks}.
\newblock \bibinfo{journal}{\emph{Proceedings of the ACM on Measurement and
  Analysis of Computing Systems}} \bibinfo{volume}{7}, \bibinfo{number}{2}
  (\bibinfo{year}{2023}), \bibinfo{pages}{1--33}.
\newblock


\bibitem[Tang et~al\mbox{.}(2023b)]%
        {tangethical}
\bibfield{author}{\bibinfo{person}{Yuzhe Tang}, \bibinfo{person}{Kai Li},
  \bibinfo{person}{Yibo Wang}, {and} \bibinfo{person}{Jiaqi Chen}.}
  \bibinfo{year}{2023}\natexlab{b}.
\newblock \showarticletitle{Ethical Challenges in Blockchain Network
  Measurement Research}. In \bibinfo{booktitle}{\emph{Workshop on Ethics in
  Computer Security (EthiCS)}}.
\newblock


\bibitem[Vasek et~al\mbox{.}(2014)]%
        {vasek2014empirical}
\bibfield{author}{\bibinfo{person}{Marie Vasek}, \bibinfo{person}{Micah
  Thornton}, {and} \bibinfo{person}{Tyler Moore}.}
  \bibinfo{year}{2014}\natexlab{}.
\newblock \showarticletitle{Empirical analysis of denial-of-service attacks in
  the bitcoin ecosystem}. In \bibinfo{booktitle}{\emph{Financial Cryptography
  and Data Security: FC 2014 Workshops, BITCOIN and WAHC 2014, Christ Church,
  Barbados, March 7, 2014, Revised Selected Papers 18}}. Springer,
  \bibinfo{pages}{57--71}.
\newblock


\bibitem[Wahrst{\"a}tter et~al\mbox{.}(2024)]%
        {wahrstatter2023blockchain}
\bibfield{author}{\bibinfo{person}{Anton Wahrst{\"a}tter},
  \bibinfo{person}{Jens Ernstberger}, \bibinfo{person}{Aviv Yaish},
  \bibinfo{person}{Liyi Zhou}, \bibinfo{person}{Kaihua Qin},
  \bibinfo{person}{Taro Tsuchiya}, \bibinfo{person}{Sebastian Steinhorst},
  \bibinfo{person}{Davor Svetinovic}, \bibinfo{person}{Nicolas Christin},
  \bibinfo{person}{Mikolaj Barczentewicz}, {et~al\mbox{.}}}
  \bibinfo{year}{2024}\natexlab{}.
\newblock \showarticletitle{Blockchain Censorship}.
\newblock \bibinfo{journal}{\emph{WWW'24: Proceedings of the ACM Web Conference
  2024, Singapore}} (\bibinfo{year}{2024}).
\newblock


\bibitem[Wahrst{\"a}tter et~al\mbox{.}(2023)]%
        {wahrstatter2023time}
\bibfield{author}{\bibinfo{person}{Anton Wahrst{\"a}tter},
  \bibinfo{person}{Liyi Zhou}, \bibinfo{person}{Kaihua Qin},
  \bibinfo{person}{Davor Svetinovic}, {and} \bibinfo{person}{Arthur Gervais}.}
  \bibinfo{year}{2023}\natexlab{}.
\newblock \showarticletitle{Time to Bribe: Measuring Block Construction
  Market}.
\newblock \bibinfo{journal}{\emph{arXiv preprint arXiv:2305.16468}}
  (\bibinfo{year}{2023}).
\newblock


\bibitem[Wang et~al\mbox{.}(2016)]%
        {wang2016abusing}
\bibfield{author}{\bibinfo{person}{Huangxin Wang}, \bibinfo{person}{Zhonghua
  Xi}, \bibinfo{person}{Fei Li}, {and} \bibinfo{person}{Songqing Chen}.}
  \bibinfo{year}{2016}\natexlab{}.
\newblock \showarticletitle{Abusing Public $\{$Third-Party$\}$ Services for
  $\{$EDoS$\}$ Attacks}. In \bibinfo{booktitle}{\emph{10th USENIX Workshop on
  Offensive Technologies (WOOT 16)}}.
\newblock


\bibitem[Wang et~al\mbox{.}(2023)]%
        {wang2023understanding}
\bibfield{author}{\bibinfo{person}{Yibo Wang}, \bibinfo{person}{Wanning Ding},
  \bibinfo{person}{Kai Li}, {and} \bibinfo{person}{Yuzhe Tang}.}
  \bibinfo{year}{2023}\natexlab{}.
\newblock \showarticletitle{Understanding ethereum mempool security under
  asymmetric dos by symbolic fuzzing}.
\newblock \bibinfo{journal}{\emph{arXiv preprint arXiv:2312.02642}}
  (\bibinfo{year}{2023}).
\newblock


\bibitem[Yaish et~al\mbox{.}(2023)]%
        {yaish2023speculative}
\bibfield{author}{\bibinfo{person}{Aviv Yaish}, \bibinfo{person}{Kaihua Qin},
  \bibinfo{person}{Liyi Zhou}, \bibinfo{person}{Aviv Zohar}, {and}
  \bibinfo{person}{Arthur Gervais}.} \bibinfo{year}{2023}\natexlab{}.
\newblock \showarticletitle{Speculative Denial-of-Service Attacks in Ethereum}.
\newblock \bibinfo{journal}{\emph{Cryptology ePrint Archive}}
  (\bibinfo{year}{2023}).
\newblock


\bibitem[Zhao et~al\mbox{.}(2024)]%
        {zhao2024dethna}
\bibfield{author}{\bibinfo{person}{Chonghe Zhao}, \bibinfo{person}{Yipeng
  Zhou}, \bibinfo{person}{Shengli Zhang}, \bibinfo{person}{Taotao Wang},
  \bibinfo{person}{Quan~Z Sheng}, {and} \bibinfo{person}{Song Guo}.}
  \bibinfo{year}{2024}\natexlab{}.
\newblock \showarticletitle{DEthna: Accurate Ethereum Network Topology
  Discovery with Marked Transactions}.
\newblock \bibinfo{journal}{\emph{arXiv preprint arXiv:2402.03881}}
  (\bibinfo{year}{2024}).
\newblock


\bibitem[Zhou et~al\mbox{.}(2021a)]%
        {zhou2021a2mm}
\bibfield{author}{\bibinfo{person}{Liyi Zhou}, \bibinfo{person}{Kaihua Qin},
  {and} \bibinfo{person}{Arthur Gervais}.} \bibinfo{year}{2021}\natexlab{a}.
\newblock \showarticletitle{A2mm: Mitigating frontrunning, transaction
  reordering and consensus instability in decentralized exchanges}.
\newblock \bibinfo{journal}{\emph{arXiv preprint arXiv:2106.07371}}
  (\bibinfo{year}{2021}).
\newblock


\bibitem[Zhou et~al\mbox{.}(2021b)]%
        {zhou2021high}
\bibfield{author}{\bibinfo{person}{Liyi Zhou}, \bibinfo{person}{Kaihua Qin},
  \bibinfo{person}{Christof~Ferreira Torres}, \bibinfo{person}{Duc~V Le}, {and}
  \bibinfo{person}{Arthur Gervais}.} \bibinfo{year}{2021}\natexlab{b}.
\newblock \showarticletitle{High-frequency trading on decentralized on-chain
  exchanges}. In \bibinfo{booktitle}{\emph{2021 IEEE Symposium on Security and
  Privacy (SP)}}. IEEE, \bibinfo{pages}{428--445}.
\newblock


\end{thebibliography}


\appendix
\section{Blockchain lookup}
\label{appendix:dropped_tx}
This section validates our definition of \textit{dropped transactions}. 
We systematically examine all transactions observed in Flashbots' dataset to ascertain their eventual inclusion on the blockchain, differentiating them from dropped transactions---those that enter the txpool but fail to be included on the blockchain. 
Our analysis involves comparing the blockchain information up to the future seven days from the observation timestamp.
There is a potential for mislabeling a transaction if it takes longer than seven days to be a part of the blockchain.
To substantiate the adequacy of the 7-day lookup window, we experiment to check the variation in our results (i.e., optimizing the extent of blockchain lookup required to define dropped transactions). 
Utilizing 100 days of data starting from September 1st, we compute the number of dropped transactions for each day, calibrating the length of the blockchain lookup by extending the reference point into the future.

In Figure~\ref{fig:blockchain_lookup}, the $x$-axis represents the number of future days of blockchain data utilized, while the $y$-axis denotes the level of dropped transactions (left: the mean dropped transactions per day, right: normalized by the amount of dropped transactions when $x=0$). 
The graph illustrates a marginal decrease in the number of dropped transactions as the window size increases. 
This trend implies that examining more than seven days of future blockchain is adequate for identifying dropped transactions. 
Specifically, extending the blockchain lookup by one additional week only results in a negligible 0.2\% change in transactions. 
If a transaction persists in the network for over one week, chances of being incorporated into the blockchain become exceedingly low.

\begin{figure}
  \begin{subfigure}{0.4\linewidth}
   \includegraphics[width=\textwidth]{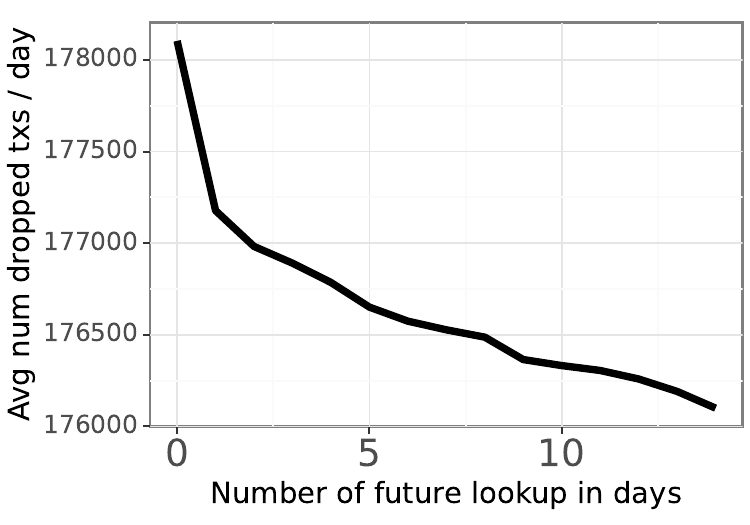}
    \caption{Mean}
  \end{subfigure}
  \begin{subfigure}{0.4\linewidth}
    \includegraphics[width=\textwidth]{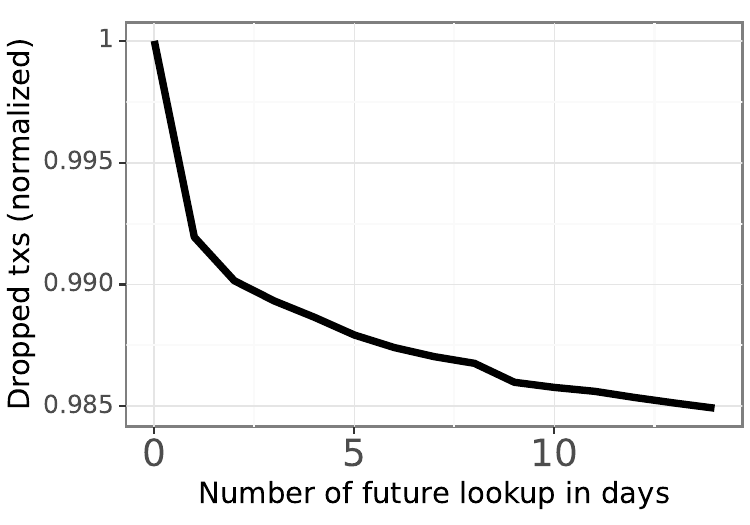}
    \caption{Normalized}
  \end{subfigure}
  \caption{Number of dropped transactions.}
  \label{fig:blockchain_lookup} 
\end{figure}

\section{Erigon propagation policy}
\label{appendix:erigon_prop_policy}
This section illustrates the change in Erigon's transaction propagation policy. 
Erigon had adopted ``square root policy'' just like Geth until v2.49.0. 
Refer to \href{https://github.com/ledgerwatch/erigon/blob/61364e8a0199c7aa2aa8f8e8377d47f6c65c2ab1/cmd/sentry/sentry/sentry_grpc_server.go#L903}{\texttt{SendMessageToRandomPeers}} used in \href{https://github.com/ledgerwatch/erigon-lib/blob/main/txpool/send.go#L73}{\texttt{BroadcastPooledTxs}}.
However, there have been two major changes to this initial implementation.

On August 23rd, 2023 (\href{https://github.com/ledgerwatch/erigon/pull/8030/commits/0974c61caeb88db25f23242f0fe0dc71b71ebd80}{\#8030}), Erigon stopped the square root policy and started to broadcast (\texttt{0x02}) to every peer the node connects to. 
The intention is to propagate the block to every node, but the change in \texttt{SendMessageToRandomPeers} function affects the transaction propagation simultaneously (after version v2.49.0). 
Next, on Dec 4th, 2023 (\href{https://github.com/ledgerwatch/erigon/pull/8742/commits/20d9e6dd70b62901eb5748203f7125b4f5845387}{\#8271}), Erigon decided to broadcast (\texttt{0x02}) to a constant number of 3 peers and announce (\texttt{0x08}) to 6 peers (after v2.55.0). 
The intention is to reduce the burden of outgoing traffic. 
eriogn's analysis~\cite{erigon_propagation_policy_analysis} illustrates that outbound traffic reduces from 5.5-6.5 MiB/s to 3-3.5 MiB/s.
Our reconstruction method (\S\ref{subsec:estimate_gx}) does not apply after v2.49.0 because an Erigon node 1) does not differentiate between \texttt{0x02} and \texttt{0x08} until v2.55.0, and 2) does not announce to all peers after v2.55.0.

\section{Manual analysis of the empirical attack}
\label{appendix:intention}
We manually look at some attackers' addresses and list two possible motivations for the attack.
The first interpretation is to disrupt the service or the network. 
If the account sends many transactions with insufficient addresses or past nonce, the attacker has no intention of making transactions on-chain, but solely to cause a disturbance to the service/network (i.e., an amplification attack). 
Figure~\ref{fig:num_spam_overtime} summarizes the number of attacks over time (on an hourly basis). 
There are more than 60 instances of attacks on November 26th between 7--8 AM (UTC). 
Many of them appear to send transactions from accounts with zero balances. 
We find similar cases on September 12th, 5 PM, October 7th, 5 PM, and November 20th, 9 PM.
The number of attacks does not appear to be correlated with the Ether price fluctuation. 

\begin{figure}[]
    \centering
    \includegraphics[width=0.75\linewidth]{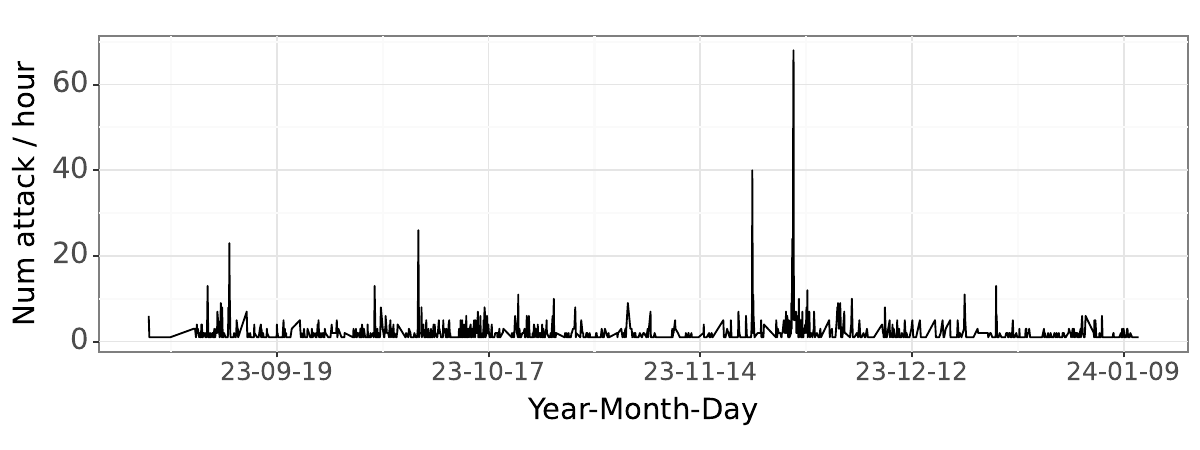}
    \caption{Number of attacks observed per hour, over time (September 1st, 2023--January 11th, 2024).}
    \label{fig:num_spam_overtime}
\end{figure}

The second interpretation is to increase the chance of transaction inclusion. 
The intention is different from our attack model.  
We look into two accounts (0x3d9e..., 0x443d...) that slightly alter the gas limit, and send thousands of invariant transactions in one block slot. 
Those accounts appear to be the arbitrage MEV bots. 
By sending thousands of transactions at once, the transactions may reach out to the nodes faster than others or fill up the txpool to prevent other bots' transactions from being executed. 
This behavior is partly discussed in another study by Zhou et al~\cite{zhou2021a2mm}. 
While many of their on-chain transactions get reverted in the middle of the execution, there are some successful complete arbitrage transactions.  
One transaction (0x3d1d...) has made 1.42 Ether by taking an arbitrage between Uni Swap and Sushi Swap, which would be 2,909 USD at the time of execution. 
As long as the cost (i.e., the amount of gas fee that the MEV bot pays for on-chain transactions and the computational cost of generating invalid transactions) does not exceed the profit of successful arbitrage, the bot has an incentive to spam the network.  

Despite different attack intentions (or scale), the attacker crafts many invalid transactions and disrupts the service/network. 
The prevalence of the attack in the current P2P network corroborates the practicality/feasibility of our proposed attack.

\section{MLE estimate of peer connections $\hat \theta_i$ and $\hat x_i$}
\label{appendix:mle_estimate}

\begin{figure*}[h]
  \begin{subfigure}{0.3\linewidth}
   \includegraphics[width=\textwidth]{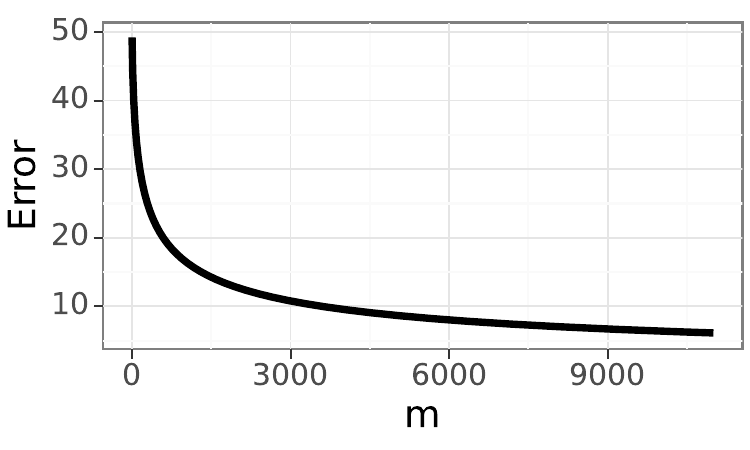}
  \end{subfigure}
  \begin{subfigure}{0.3\linewidth}
    \includegraphics[width=\textwidth]{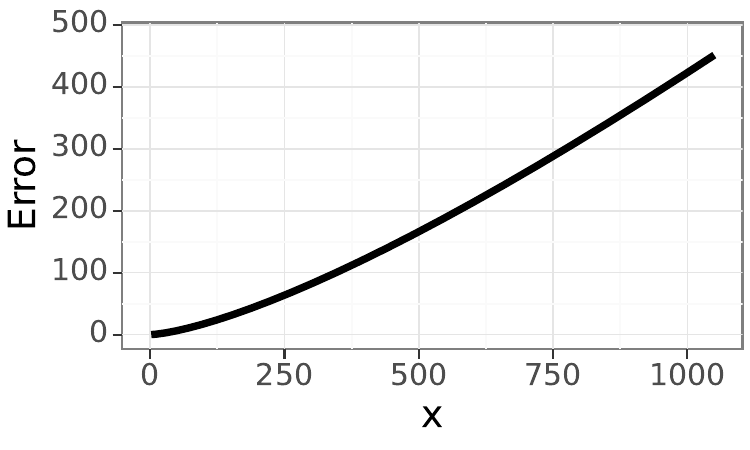}
  \end{subfigure}
  \begin{subfigure}{0.3\linewidth}
    \includegraphics[width=\textwidth]{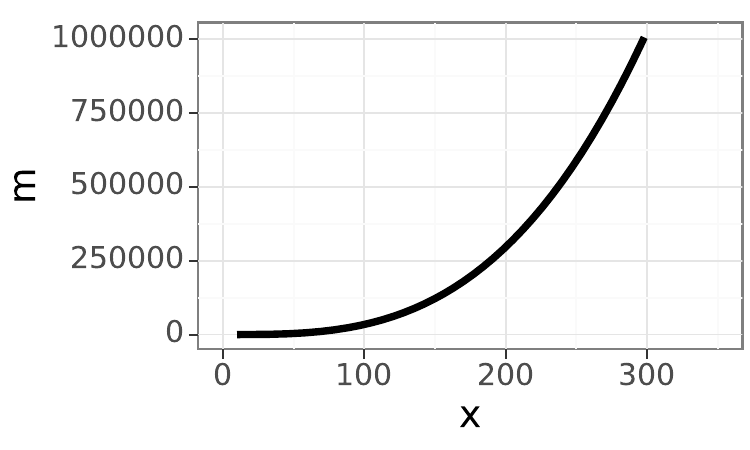}
  \end{subfigure}
  \caption{The relationship among $m$, $x_i$, and the amount of error.}
  \label{fig:mle_error_geth} 
\end{figure*}

This section complements the mathematical formulation presented in \S\ref{subsec:estimate_gx}.
From the likelihood defined in Eqn.~(\ref{eq:likelihood}), log-likelihood function is 

$$
\ln l(\theta_i,  m_2, m_2 + m_8) = \ln {m_2 + m_8 \choose m_2} + m_2 \ln \theta_i- m_8 \ln (1-\theta_i) \ .
$$
We take the derivative of this function w.r.t $\theta_i$ and set the equation to 0 to maximize the likelihood, which produces the unbiased parameter $\hat \theta_i$ (i.e., the ratio of \texttt{0x02} messages).

\begin{small}
$$
  \frac{d \ln l(\theta_i, m_2, m_2 + m_8)}{d \theta_i} 
= \frac{m_2}{\theta_i} + \frac{m_8}{1-\theta_i}  
= 0 \Leftrightarrow
\hat \theta_i = \frac{m_2}{m_2 + m_8}  \ .
$$
\end{small}
This allows reconstruction in Eqn.(\ref{eq:sqrt_policy_geth}).

We next calculate the variance of $\hat \theta_i$ to quantify the level of uncertainty for the estimate. 
Assuming that the MLE estimator is asymptotically normally distributed (i.e., when the sample size is large enough), the estimator $\hat \theta_i$ is normally distributed. 
Based on the Fisher information, the variance ($Var[\hat \theta]$) can be derived by calculating the 2nd derivative on the log-likelihood w.r.t. $\theta_i$. 

\begin{align*}
& \frac{d^2 \ln l(\theta,  m_2, m_2 + m_8)}{d \theta_{i}^2} \\
& =  - \frac{m_2}{\theta^2} + \frac{m_8}{(1-\theta)^2} \quad \text{Using $\hat \theta_{i}= \frac{m_2}{m_2 + m_8}$} \\
& = - \frac{\hat \theta_{i}(m_2+ m_8)}{\hat \theta_{i}^2} + \frac{(1 - \hat \theta_{i}) (m_2 + m_8)}{(1-\hat \theta_{i})^2} 
= - \frac{m}{\hat \theta_{i}( 1 - \hat \theta_{i})} \ .
\end{align*}

The variance is the negative inverse of the 2nd derivative; thus $Var[\hat \theta_i] = \frac{\hat \theta_i(1 - \hat \theta_i)}{m}$. 
It is at most (bounded by) $\frac{1}{2\sqrt{m}}$ when $\hat \theta=\frac{1}{2}$ (maximum). 
Within two standard deviation (std) $\sigma_{\hat \theta_i}$ (which covers around 95\%), a confidence interval (CI) of $\hat \theta_i$ is $\hat \theta_i \pm 2\sigma_{\hat \theta_i} = \hat \theta_i + \frac{1}{\sqrt{m}}$. 
However, after the reconstruction of $\hat x_i$ using Eqn.(\ref{eq:sqrt_policy_geth}), the variance of $\hat x_i$ does not appear to have a closed-form solution, so we use heuristics to determine the credibility of our estimate $\hat x_i$. 
In particular, since the variance of $\hat x_i$ would expand more quickly when $\hat \theta_i$ is small, we exclude the uncertain estimates based on the level of $\hat \theta_i$ and $Var[\hat \theta_i]$.
We calculate $\hat x_i$ from two $\hat \theta$s: 1) $\hat \theta_i$, and 2) $\hat \theta_i + (1/\sqrt{m})$ (two std apart), and refer to the difference of the two points as an ``error,'' $\varepsilon$. 
If the error is more than 10 (i.e., the estimated number of peers $\hat x_i$ is likely to deviate more than 10 peers), we exclude the estimate. 

\begin{equation}
\varepsilon =   \underbrace{\left(\frac{1}{\hat \theta_i}\right)^2}_{\theta_i = \hat \theta_i} - \underbrace{\left(\frac{1}{\hat \theta_i + (1/\sqrt{m})}\right)^2}_{\theta_i = \hat \theta_i + (1/\sqrt{m})}  = x_i -\left(\frac{1}{\sqrt{x}} + \frac{1}{\sqrt{m}}\right)^{-2} \ .
\end{equation} 

We next look at the relationship between $x$, $m$, and the error $\varepsilon$.
Figure~\ref{fig:mle_error_geth} (left/middle) illustrates the change in the level of error by increasing $m$ and $x$. 
When the number of observations $m$ increases, the error marginally decreases. 
When the peer connection $x$ increases, the error slowly increases. 
Figure~\ref{fig:mle_error_geth} (right) shows the relationship between $x$ and $m$ while fixing the amount of error.
A much larger number of samples is necessary for the node with many connections to obtain a reliable estimate.

\section{Customized nodes}
\label{subsec:customized_nodes}

\begin{figure}[]
  \includegraphics[width=0.7\linewidth]{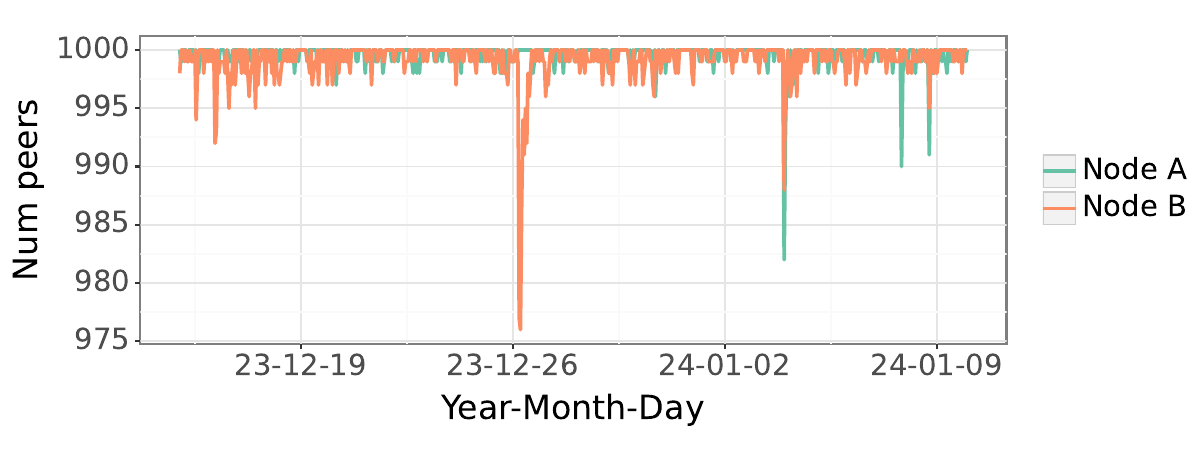}
  \caption{Number active peer connections in our nodes' connections. The $y$-axis does not start at 0.}
  \label{fig:num_peers} 
\end{figure}

\begin{figure}[]
  \includegraphics[width=0.7\linewidth]{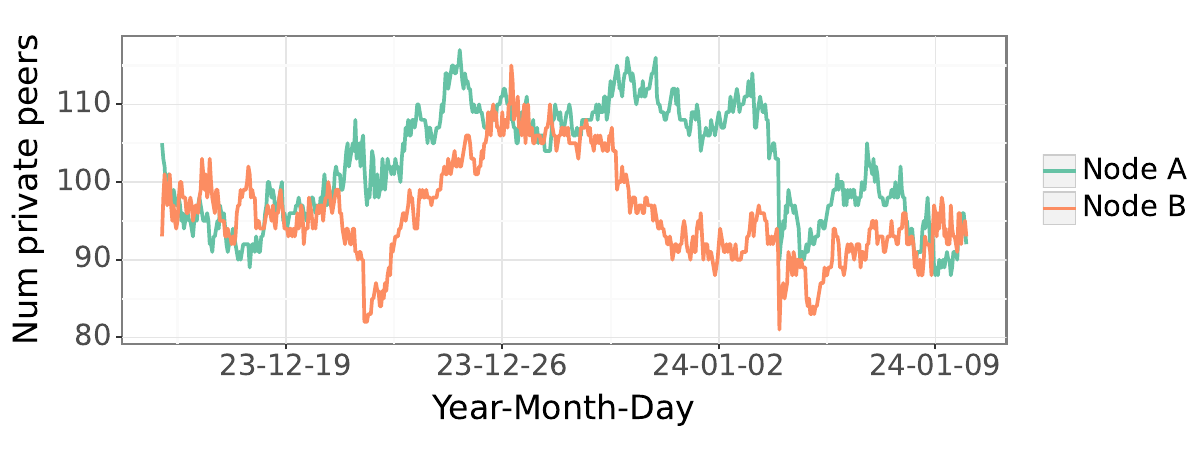} 
  \caption{Number active customized peer connections in our nodes' connections.}
  \label{fig:num_private_peers} 
\end{figure}

We find a cluster of nodes that customize the existing clients and collectively develop the software privately. 
As noted in \S\ref{subsec:private_data}, we collect the node name to profile P2P network nodes.
Node names (e.g., Geth/v1.13.4-stable-3f907d6a/linux-amd64/go1.21.3) generally follow a convention indicating 1) 
software name, 2) version/branch and the first 4 bytes of the last \texttt{git commit} hash,
3) operating system, and 4) programming language and its version. 
We use regular expressions to extract the version and \texttt{git commit} hash. 
If the node owner forks existing projects, modifies any part of the code, and updates the repository, a different \texttt{git} hash is generated.
To determine whether the node is customized, we call the Github API\footnote{\url{https://docs.github.com/en/rest/commits/commits?apiVersion=2022-11-28}} to check whether the commit publicly exists in the corresponding software repository. 
We call nodes with non-public \texttt{git commit} hashes \emph{customized nodes} (not modified nodes).
Given the space of possible hashes (4 bytes: $16^8$), the probability of generating two identical hashes by mere coincidence is extremely low, which suggests the cluster of nodes being operated by the same entity. 

We identify many clusters of customized nodes. 
For example, the largest cluster has 250 nodes, and all reside in the same ISP (data center) in one (US) city.
There are also two clusters (200+ different IDs) that appear to be operated by the same entity, but those nodes are spread out across multiple ISPs and 19 cities. 
We infer that they may belong to one relay service that attempts to reduce the latency by deploying nodes globally. 
Figure~\ref{fig:num_peers} illustrates the number of active connections in our two deployed nodes (based on the timestamp of connection/disconnection for each peer), which shows the stability of our nodes over time.  Figure~\ref{fig:num_private_peers} depicts the number of customized nodes; around 100 (10\%) are customized nodes.

\section{Software client, ISP}
\label{appendix:software_isp}
We investigate software client market share to validate our choice
of transaction propagation policies in \S\ref{subsec:network_waste}.
To exclude unstable/non-functioning nodes, we produce statistics
from nodes that maintain peer connections to our two nodes. Our
reported statistics are thus different from those at websites such as
Etherscan~\cite{etherscan}, which include all peers from other chains
in the P2P network. We observe a software market share that remains
stable over time. Geth, Erigon, and Nethermind have 75\%, 20\%, and 5\%,
respectively. We can thus focus on Geth and Erigon in our model.

We next look at the number of nodes in each cloud service, which helps
calculate the traffic costs of the attacker/modified nodes. Our data
shows the concentration of nodes in a handful of autonomous systems
(AS). Deploying the blockchain node in the cloud is expensive, but
brings potential latency reduction benefits by co-locating with other
nodes in the same data center. The top four ASs are all cloud services
and account for more than 50\% of the active nodes. In particular, 20\%
of all active nodes appear to be hosted on AWS.
\section{Attack scenario}
\label{appendix:attack_scinario}
We aim to estimate the highest possible traffic costs that modified nodes would incur when an attacker saturates their outgoing traffic with its invalid transactions. 
The attacker can utilize at most 400 Ethereum blockchain accounts (as shown in Figure~\ref{fig:attack_simulation}) to evict existing transactions in the txpool and keep refreshing it with the new invalid ones, causing the modified nodes to continuously forward invalid transactions to their neighbors.
The attacker can also send
transactions from multiple nodes with different IPs, and distribute
the attack to multiple modified nodes to scale up the attack until the
bandwidth limit (i.e., \textit{Distributed EDoS attack}). 
This helps
attackers circumvent any network management tools deployed by modified
nodes (e.g., anomaly detection---one of the solutions to classical
EDoS attacks). We partly reference Kumar's work~\cite{kumar2007smurf}
to calculate the total link capacity of intermediate nodes (i.e.,
modified nodes in our case). Table~\ref{tab:attack_multiple}
summarizes the assumptions and the final cost of modified nodes.
We use the bandwidth of AWS EC2 instances for attacker/modified
nodes and the minimum bandwidth requirement by Geth for the regular
nodes.\footnote{\url{https://geth.ethereum.org/docs/getting-started/hardware-requirements}} In our case, the scale of the attack (the amount of
waste) is mainly constrained by the number/bandwidth of modified/regular
nodes since the attacker's traffic gets easily amplified by TAF. Based
on our calculation, each modified node could spend at most 88,904 USD, 
or roughly 8M USD per month in aggregate 
if all the modified nodes are from the
same entity.

\begin{table}[]
\centering
\scalebox{1}{
\begin{tabular}{@{}lllll@{}}
\toprule
              & Capacity per node & \# of nodes       & Capacity (network)          & Cost          \\ \midrule
$\mathcal{M}$ & $\beta_{\mathcal{M}, out}$    & $ \gamma N$       & $ \beta_{\mathcal{M}, out} \gamma N$      &               \\
              & 2.5~GB/second  & 90              & 225~GB/second               & \$89K/node  \\
              & 6480~TB/month  &                   & 583,200~TB/month            & \$8M/all \\ \midrule
$\mathcal{R}$ & $\beta_{\mathcal{R}, in}$     & $ (1 - \gamma) N$ & $ \beta_{\mathcal{R}, in} (1 - \gamma) N$ &               \\
              & 12.5~MB/second & 5910               & 73.88~GB/second             &              \\
              & 32.4~TB/month  &                   & 191,484~TB/month            &               \\ \bottomrule
\end{tabular}
}
\caption{One attack scenario ($\mathcal{M}$: modified node, $\mathcal{R}$: regular node) with link capacity constraints.}
\label{tab:attack_multiple}
\end{table}

Alternatively, the attacker might opt to perform this attack
moderately over an extended duration to evade detection, rather
than maximizing traffic volume over a brief period. Previous
literature~\cite{chowdhury2017economic} highlights the potential
``stealthy nature'' of the EDoS attack (compared to DoS) where the
attacker carefully chooses the level of traffic under the detection
threshold set by the victim. 

This strategy is indeed effective under the AWS's pay-as-you-go pricing model. 
AWS charges 90 USD/TB until 10TB, 85 USD/TB for the next 40TB, 70 USD/TB for the next 100TB, and 50 USD greater than 150TB. 
Figure~\ref{fig:eaf} illustrates the decrease in EAF ($y$-axis) based on the amount of outgoing traffic ($x$-axis) due to AWS's dynamic pricing. 
The attacker's economic benefit marginally decreases over the scale of the attack.  
It might be more economically effective to control the level of the attack.

\begin{figure}[]
    \centering
    \includegraphics[width=0.65\linewidth]{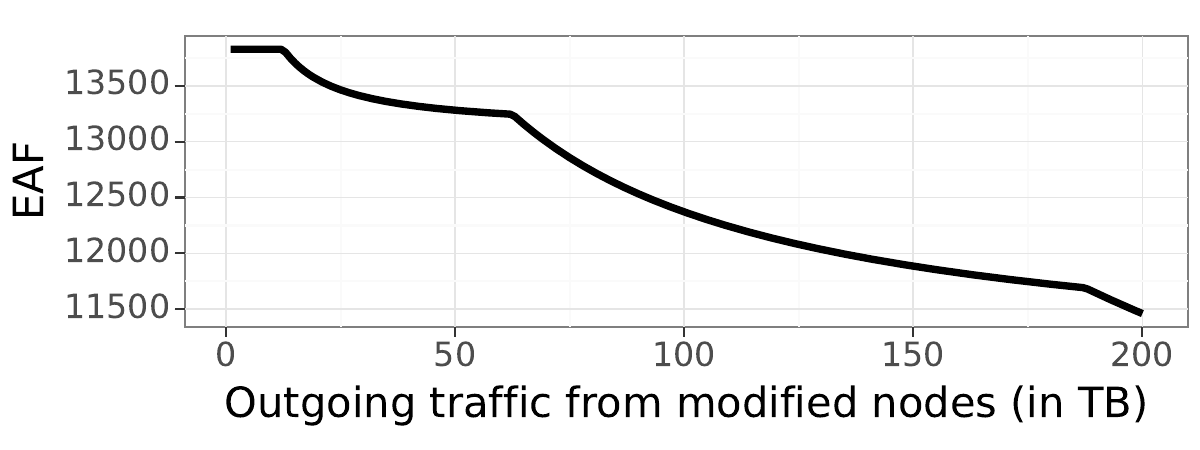}
    \caption{EAF based on the traffic usage of a modified node.}
    \label{fig:eaf}
\end{figure}
\section{Analysis of validation simulation}
\label{appendix:cpu_profile}
This supplementary section explains why state-dependent checks consume the majority of the processing time as illustrated in \S\ref{subsec:val_simulation}. Figure~\ref{fig:pprof} dissects the distribution of the CPU processing time. 

Ethereum uses the MPT to maintain the blockchain state, which is a tree structure where a leaf node stores the value of persistent data (e.g., account balance), and all intermediate nodes in the path from the root node to the leaf node correspond to the key of the data's value (e.g., account address). 
Fetching account nonces also requires multiple lookups in the tree. 
This traversal process is time-consuming for several reasons~\cite{he2024nurgle}. 
First, due to the large amount of state, the tree can become deep, increasing the number of nodes that must be traversed to find a particular piece of data. 
Second, disk access plays a significant role because the MPT structure often resides on disk. 
Slow disk accesses occur when the nodes required for a lookup are not in memory and must be fetched from disk.

Although the nonce check appears to be the most expensive operation,
it is not because other state-based validations (e.g., balances) are
cheap; they appear faster primarily because when fetching the nonce,
the balances of the account are also cached in one look-up, minimizing
additional overhead (e.g., \texttt{trie.(*StateTrie).GetAccount}
in Figure~\ref{fig:pprof}). This suggests that omitting one check
while retaining another (i.e., merely checking the nonce but not the
balance) does not alter the validation process, thereby not contributing
meaningfully to latency reduction.

\begin{figure}[]
    \centering
    \includegraphics[width=0.68\linewidth]{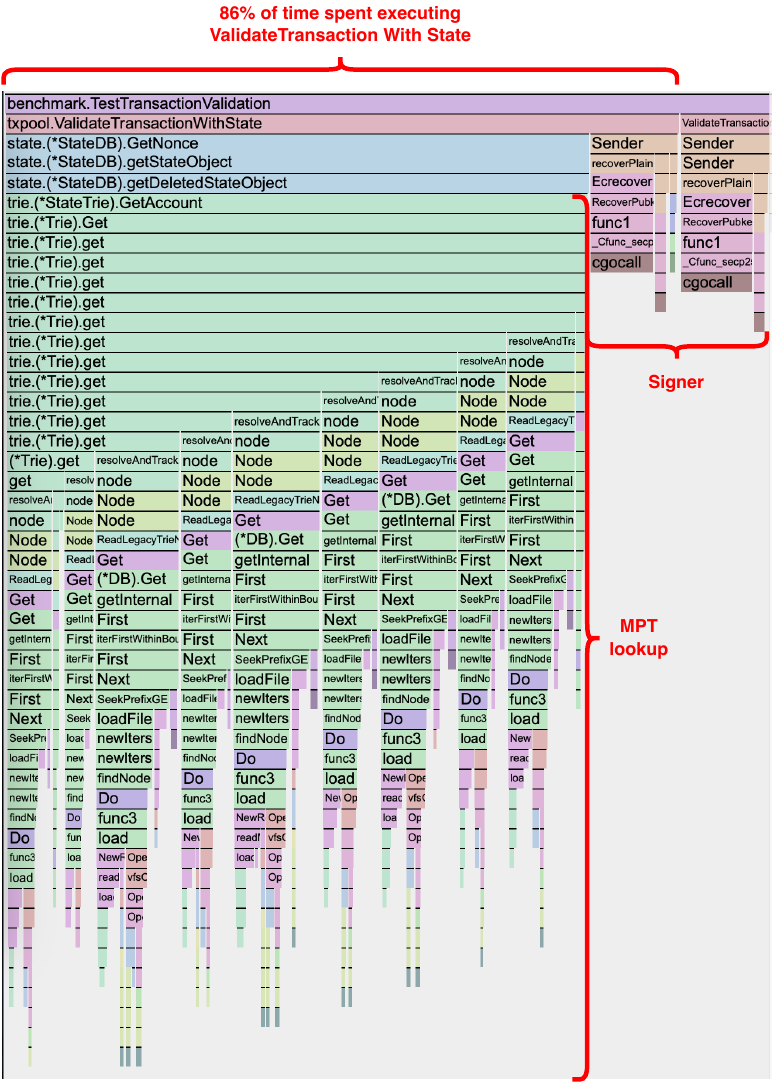}
    \caption{Profiling execution time for transaction validation --- analysis of $136,437$ Transactions from $1,000$ consecutive blocks starting at block $18,140,000$ using Golang's ``pprof.''}
    \label{fig:pprof}
\end{figure}
\section{Discussions on measuring latency on testnet}
\label{appendix:measuring_latency}
In \S\ref{subsec:run_modified}, we measure the latency (confirmation time) between the regular and the modified node at the test net Holesky. 
This section discusses how we process the measurement results and explains the
limitations.

We observe that users send a large number of transactions in a short period (e.g., attacks and tests), particularly due to the absence of 
transaction fees in testnet. 
To eliminate the effect of outliers, we count the number of transactions for five-minute intervals and exclude the period when the number of observed transactions is more than 3 standard deviations above the mean---excluding 1.14\% of our total observation time.

Even after excluding abnormal activities, the market conditions differ significantly over time.
Figure~\ref{fig:confirmation_time_holesky_daily} shows the median confirmation time for each day. 
The colors represent the type of node running.
It is clear that the transaction inclusion time varies by a margin of a few seconds each day, which could offset the latency reduction benefits gained from skipping validations. 

In addition, the connected peers differ as well. Our node connects
to a different set of peers even with the same public key (around
15\% overlap), indicating that testnet peers might change very 
frequently. 
Furthermore, the testnet is much less active than the mainnet (around 30\%). 
If the node receives many transactions at once, it may not be able to process transactions simultaneously, which adds latency to the regular node's processing time. 
Finally, while this experiment only evaluates a single modified node, deploying or replaying multiple nodes without validation could lead to a cumulative latency reduction. 

\begin{figure}[]
    \centering
    \includegraphics[width=0.7\linewidth]{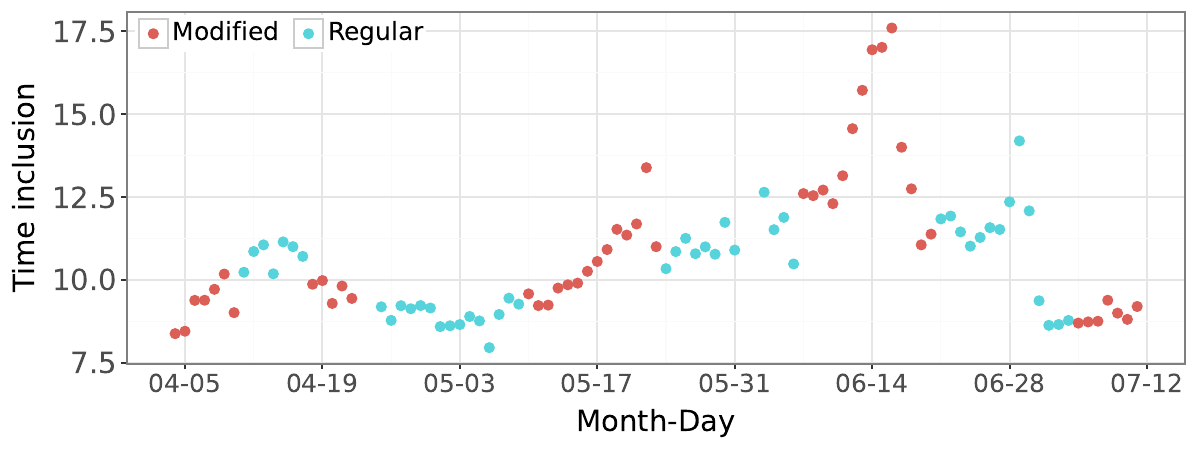}
    \caption{Time inclusion for a modified (red) and regular node (blue) on a daily basis (median)}
    \label{fig:confirmation_time_holesky_daily}
\end{figure}

\section{Time to Money Conversion}
\label{appendix:time_to_money}
This section shows how we estimate the economic value of latency in \S\ref{subsec:cost_benefit_analysis}.  
Wahrst{\"a}tter et     
al~\cite{wahrstatter2023time} empirically measure the increase in the Ethereum's block bid value (in one slot). 
In particular, they derive a polynomial function $P(x) =      
-1.99 x^3 + 2.44 x^2 + 32.5 x + 40.77$, where $x$ denotes a block       
slot time in \textit{seconds} at which a bid is submitted, and $P(x)$   
represents the profit from submitting the bid at time $x$, normalized   
by the average bid value at $x=-2$. 
The derivative $P'(x)$ thus represents the relative profit increase at time $x$. 
In other words, the time $x$ that maximizes $P'(x)$ is     
the sweetest spot in reducing latency, which is $P''(x) = 0$ at $x=0.409$s.
We estimate the millisecond profit increase ($P(x)-P(x-0.001)$) at $t=0.409, 1, 2, 2.5$ in Table~\ref{tab:latency_vs_profit}. 
Assuming an average MEV bid of 0.06
ETH\footnote{https://mevboost.pics} at an ETH price of \$2,500, the expected profit increase is 0.00002 ETH (\$0.050) per millisecond at maximum.

\begin{table}[]
    \centering
    \scalebox{0.9}{
    \begin{tabular}{lllll}
    \toprule
       Submit the block at & $0.409s$ (max) & $1s$ & $2s$  & $2.5s$ \\ \midrule
       \multicolumn{5}{c}{For Every 1ms Latency Reduction} \\ \midrule
       \% Profit Increase & $0.034\%$ & $0.031\%$ & $0.018\%$ & $0.0074\%$ \\
       Profit Increase (ETH) & $\approx$ 0.000020 & $\approx$ 0.000019 & $\approx$ 0.000011 & $\approx$ 0.0000044\\
       Profit Increase (USD) & $\approx$ \$0.050  & $\approx$ \$0.047  & $\approx$ \$0.028 & $\approx$ \$0.011 \\ 
    \bottomrule
    \end{tabular}
    }
    \caption{Estimation of profit increase vs. latency reduction per block in Ethereum based on Wahrst\"{a}tter et al.~\cite{wahrstatter2023time}. Calculations assume an ETH price of \$2,500, and an average MEV bid of 0.06 ETH.}
    \label{tab:latency_vs_profit}
\end{table}
\end{document}